\documentclass[a4paper, amsfonts, amssymb, amsmath, reprint, showkeys, nofootinbib, twoside,floatfix]{revtex4-2}
\usepackage[english]{babel}
\usepackage[utf8]{inputenc}
\usepackage{array}
\usepackage{textgreek}
\usepackage{amsthm}
\usepackage{mathtools}
\usepackage{physics}
\usepackage{xcolor}
\usepackage{graphicx}
\usepackage{textgreek}
\usepackage{upgreek}
\usepackage[left=13mm,right=13mm,top=25mm,bottom=25mm,columnsep=15pt]{geometry} 
\usepackage{adjustbox}
\usepackage{placeins}
\usepackage[T1]{fontenc}
\usepackage{lipsum}
\usepackage{csquotes}

\usepackage{fancyhdr} 
\fancyheadoffset{0cm}

\fancypagestyle{plain}{%
  \fancyhf{}%
  \fancyhead[R]{\thepage}%
}

\usepackage{xr-hyper} 
\usepackage[pdftex, pdftitle={Article}, pdfauthor={Author}]{hyperref}
\usepackage{epstopdf}
\usepackage{siunitx}
\usepackage{threeparttable}
\usepackage{tabularx}
\begin{document}
\raggedbottom

\title{Demonstration and Non-volatile Trimming of a Highly-Parallel, \\ High-Capacity Silicon Microdisk Transmitter}
\author{Chao Luan$^{1*}$}
\author{Alexander Sludds$^{1}$}
\author{Chao Li$^{1}$}
\author{Ian Christen$^{1}$}
\author{Ryan Hamerly$^{1,2}$}
\author{Dirk Englund$^{1}$}

\affiliation{$^{1}$Research Laboratory of Electronics, MIT, Cambridge, MA, 02139, USA}
\affiliation{$^{2}$NTT Research Inc., PHI Laboratories, 940 Stewart Drive, Sunnyvale, CA, 94085, USA}

\begin{abstract} 
Optical interconnects are the most promising solution to address the data-movement bottleneck in data centers. Silicon microdisks, benefiting from their compact footprint, low energy consumption, and wavelength division multiplexing (WDM) capability, have emerged as an attractive and scalable platform for optical modulation. However, microdisk resonators inherently exhibit low fabrication error tolerance, limiting their practical deployment. Here, utilizing a CMOS photonics platform, we demonstrate 1.28 Tb/s of off-die bandwidth through a 64 microdisk modulator system. In addition, we develop an automated, close-looped, non-reversible, low-loss, and picometer-precision permanent wavelength tuning technique using laser trimming, and achieve a fully passive, 5-channel dense wavelength division multiplexing (DWDM, 50 GHz spacing) transmitter. The integration of the high speed (1.28 Tb/s), low energy consumption (29 fJ/bit) and the permanent wavelength trimming lays a robust foundation for next-generation optical interconnect systems, poised to facilitate scaling of future AI and computing hardware.

\end{abstract}
\maketitle

\section*{Introduction}
The exponential growth of artificial intelligence (AI) and deep learning applications drives computational workloads from individual devices to server racks in data centers, placing unprecedented requirements on the interconnect and communication infrastructures, which need to be high-speed and low-energy-consumption~\cite{867687, 7805240, cheng2018recent, wang2023extremely, van2014trends, frank2006optimizing, waldrop2016more, keckler2011gpus}. Traditional intra- and inter- chip interconnects, which rely on electrical metal-wire links, are struggling to satisfy the increasing bandwidth and energy consumption requirements imposed by current data-hungry AI applications~\cite{cheng2018recent, lee2022beyond, 7805240}. As AI workloads scale, multi-node, multi-processor systems demand significantly much higher-speed and more energy-efficient XPU-to-XPU data communications to effectively handle large-scale computational-intensive tasks~\cite{lee2022beyond,markov2014limits, baliga2011energy,semiconductor2005international}. In contrast, optical interconnect technologies that leverage silicon photonics (SiPh) offer bandwidth and energy efficiency advantages over conventional electrical metal-wire links~\cite{hochberg2010towards,tucker2009evolution, agrell2016roadmap,rizzo2023massively, daudlin2025three, yuan20245, rizzo2022petabit, wade2018bandwidth, sun2020teraphy, sun2015single, atabaki2018integrating, wade2021error,beausoleil2012photonic, miller1997physical}. SiPh components, which are based on the total internal reflection of silicon-on-insulator (SOI) waveguides, have several distinct advantages, including compatibility with mature complementary metal–oxide–semiconductor (CMOS) manufacturing process, large bandwidth (10's THz), compact footprint ($~\upmu\text{m}$ scale), low loss (dB/cm to dB/m), and intrinsic high parallelism~\cite{timurdogan2014ultralow, shekhar2024roadmapping, rahim2021taking, li2016scaling, jalali2006silicon, soref2007past, siew2021review}. These combined advantages make SiPh interconnects an essential infrastructure components to address the large-bandwidth and low-energy-consumption requirements for contemporary data center.

For short-reach interconnects and communications, the low-complexity and low-cost intensity-modulated direct-detection (IM-DD) system remains dominant within the data centers~\cite{geravand2025ultrafast, rizzo2023massively, daudlin2025three}, where integrated electro-optic intensity modulators (EOIMs) convert the XPU electrical signals to analog optical intensities. The high-index contrast of the SiPh platform enables the realization of ultra-compact vertical-type \textit{p–n} junction microdisk EOIMs. The vertical junction microdisk modulators provide large overlap between carriers and the optical mode, in addition, by leveraging the intrinsic wavelength selectivity nature, WDM and spatial division ultiplexing (SDM) can be implemented to provide a scalable high parallel modulation across a broadband wavelength range.  Consequently, SiPh microdisk EOIMs enables large-scale, high bandwidth, high efficiency, and low energy consumption modulation in a compact footprint~\cite{rahim2021taking, timurdogan2014ultralow, rizzo2022petabit,daudlin2025three, bogaerts2012silicon}. The CMOS compatibility of SiPh also allows high-yield mass production, which is desired for low-cost optical interconnects and communications in data centers.

\begin{figure*}[t!]
\centering
\includegraphics[width=0.5\textwidth]{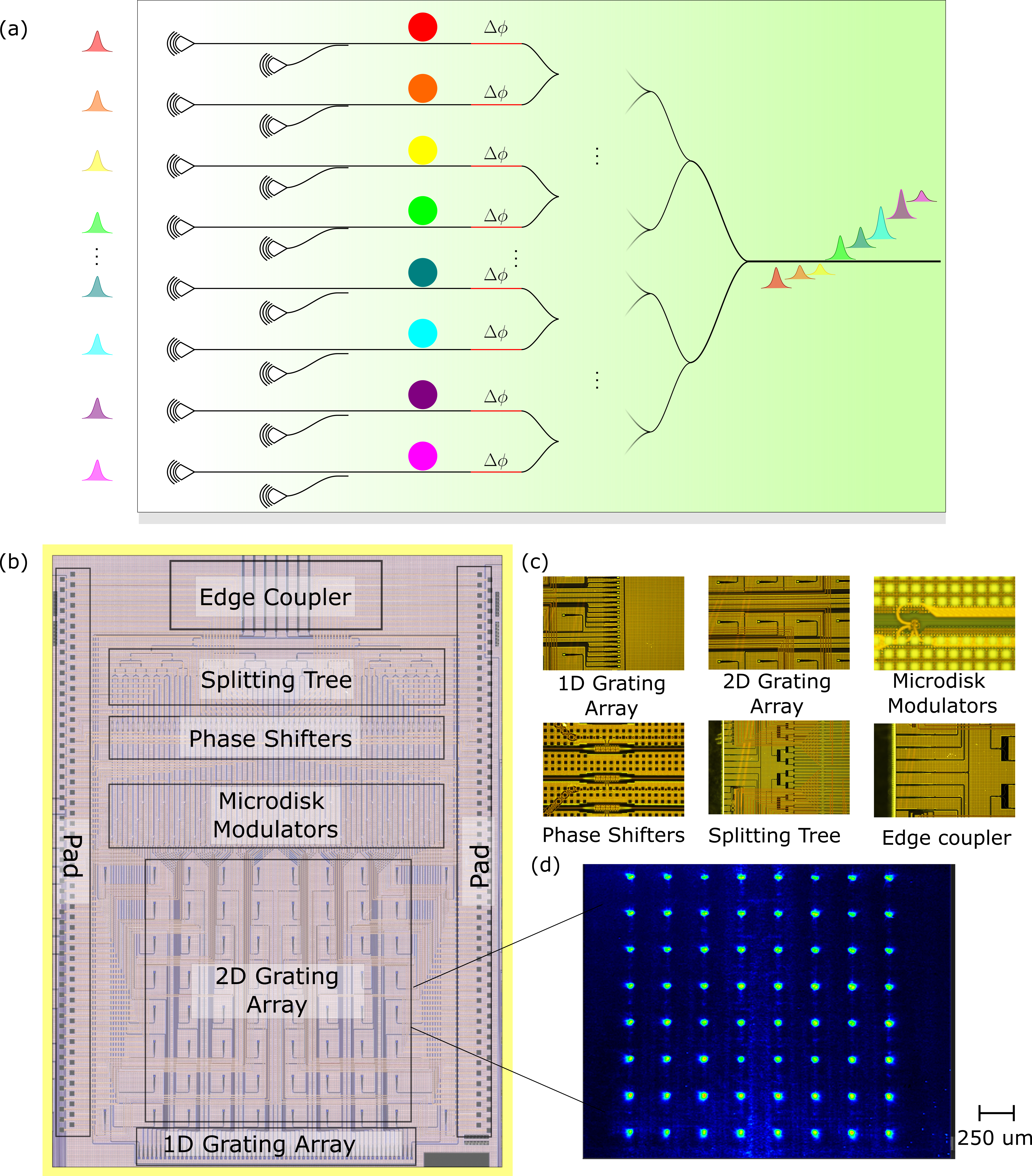}
\caption{Artistic vision and microscope image of the high parallelism, high capacity micro-disk transmitter system. (a) Schematic diagram of the 64-channel high parallelism, high capacity microdisk transmitter. The transmitter supports high parallelism SDM and allows for high per-fiber bandwidth WDM transmission. (b) Optical microscope image of the transmitter chip, from input to output, the chip includes the input grating coupler array, microdisk modulators, phase shifters, y-splitter combiner tree, and output edge coupler. (c) High resolution microscope images of the main components of the chip. (d) 2D emission image of the chip shows an 8 by 8 emitter array.}
\label{fig:1}
\end{figure*}

\begin{figure*}[t!]
\centering
\includegraphics[width=0.9\textwidth]{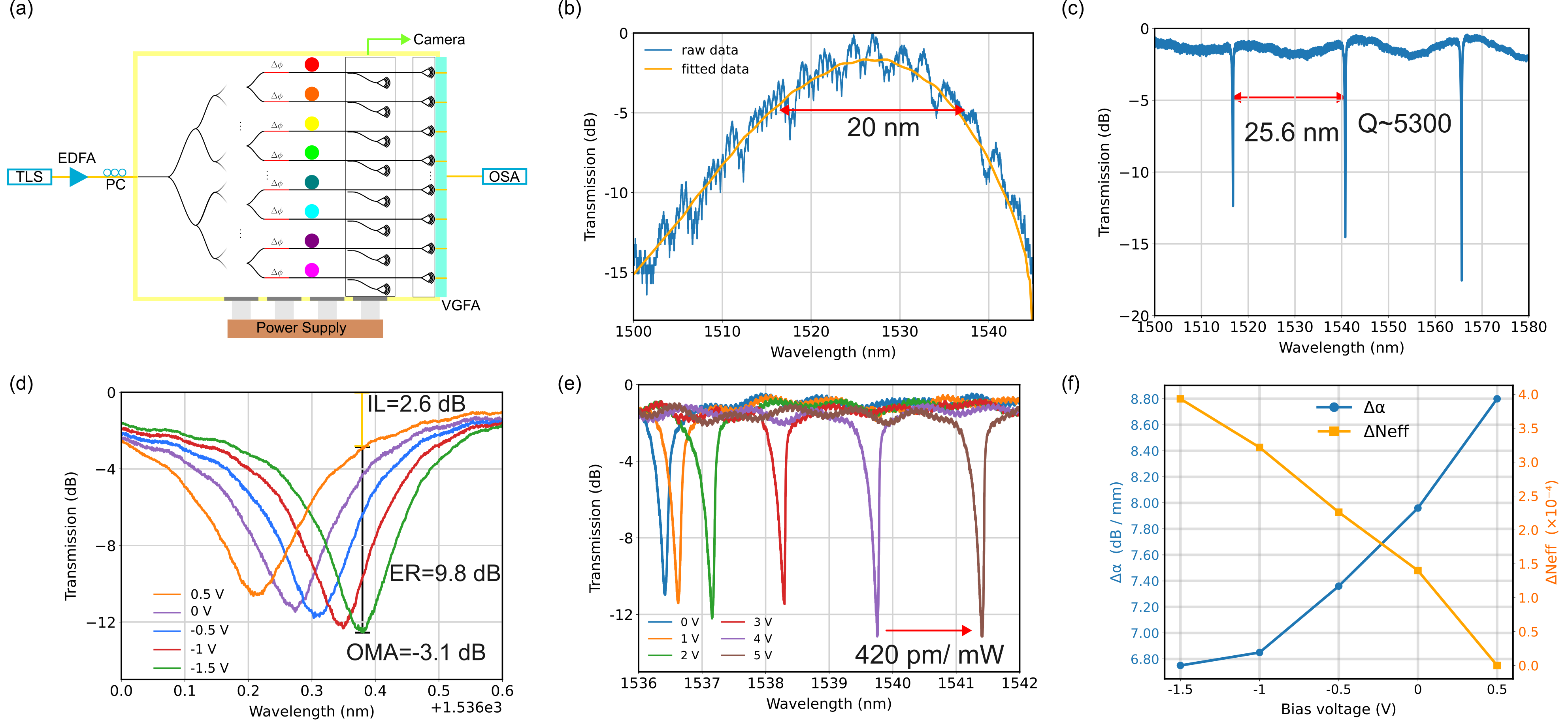}
\caption{Direct current characterization of the Microdisk Transmitter chip. (a) Experimental setup of the microdisk transmitter direct current characterization, the TLS and OSA are synchronized to provide high precision characterization, an IR camera is employed to provide parallel characterization. TLS, tunable laser; OSA, optical spectrum analyzer; PC, polarization controller; VGFA, V-groove fiber array; EDFA, Er-doped fiber amplifier. (b) Measured and fitted grating coupler transmission spectrum of the transmitter. (c) Measured broadband transmission spectrum of a single-channel transmitter at 0-V D.C voltage. (d) Measured transmission spectrum versus reversed junction voltage from 0.5 V to -1.5 V, under a 2-V voltage swing at 1536.38 nm, the modulator insertion loss (IL) is 2.6 dB, the extinction ratio (ER) is 9.8 dB, and the optical modulation amplitude (OMA) is -3.2 dB relative to the input optical power, these parameter values are nearly constant among these 64-channel microdisk modulators. (e) Measured transmission spectrum versus heater voltage ranging from 0 V to 5 V, showing a wavelength red shift tunning efficiency of 420 pm/mW. (f) Extracted absorption coefficient and refractive index change of the microdisk transmitter versus junction voltages from 0.5 V to -1.5 V.}
\label{fig:2}
\end{figure*}

While microdisk EOIMs offer compact footprints, they exhibit extreme sensitivity to fabrication errors and uncertainties~\cite{bogaerts2012silicon, nikdast2016chip, pintus2019pwm, jayatilleka2021post}. Even sub-nanometer variations in waveguide dimensions can significantly shift the spectral response of these devices. Both active tuning and post-fabrication trimming methods are employed to correct the resonance uncertainties~\cite{zhou2009athermalizing, jayatilleka2021post, schrauwen2008trimming, chu1999wavelength}. Active tuning, commonly achieved through the thermo-optic effect, though effective, it needs continuously large power consumption, posing a serious challenge to the scalability of large-scale implementation. In contrast, post-fabrication trimming permanently modify the effective index of the optical mode and doesn't need continuous power consumption. Established techniques include selective post-fabrication etching, deposition of phase change materials (PCMs), anodic oxidation, laser and thermal annealing~\cite{jayatilleka2021post, wu2025lossless, chen2025thermally, farmakidis2023scalable}. However, in standard foundry‑fabricated compact microdisk modulators the waveguides are buried beneath thick dielectric top claddings and criss‑crossed by embedded metal interconnects, thus less compatible with PCM integration, in addition, incorporate active \textit{p–n} junctions cannot tolerate the elevated post‑process temperatures required by some trimming schemes. Instead, adopt a method to locally densify the oxide cladding of each modulator via laser annealing as a practical post-fabrication trimming technique~\cite{Menssen:23}, we observe that this process increases the mode effective refractive index of the microdisk EOIM, yielding a red shift in the resonant wavelength. 

In this work, we demonstrate and perform non-volatile trimming of a highly parallel, high capacity EOIM transmitter featuring 64 compact microdisks integrated onto a single SiPh chip. The microdisk EOIM has an electro-optic bandwidth from 19 GHz to 28 GHz, an extinction ratio (ER) of 9.8-dB, an insertion loss (IL) of 2.6-dB, and an optical modulation amplitude (OMA) of -3.2 dB relative to the input optical power at a 2-V voltage swing. By implementing the transmitter in a high-speed optical transmission link, we demonstrates an on-off keying (OOK) data transmission at 20 Gbt/s (setup limited), yielding a total transmitter capacity of 1.28 Tbt/s and a low dynamic modulation energy consumption of 29 fJ/bit. Furthermore, we utilize a home-built laser annealing system to implement automated in-situ, precise, and non-reversible post-fabrication trimming of the transmitter, and achieves a wavelength red shift from 100 pm to 380 pm. By employing this non-volatile laser trimming technique, we experimentally demonstrate a thermal-tuning free, fully passive 5-channel DWDM (50 GHz channel spacing) silicon microdisk link. Compared with the untrimmed microdisk EOIM, the developed trimming technique reduces 33 \% consumption of the energy required to tune the resonant wavelength. Our non-volatile trimming approach highlights the potential of SiPh in leveraging light as a high-bandwidth and energy-efficient inter-chip communication medium, paving the way to addressing the critical scaling challenges for data-movement within modern data centers.

\section*{Architecture of the High Parallelism Microdisk Transmitter}


Figure~\ref{fig:1} illustrates the structure of the transmitter chip. 64 grating couplers, each addressed by a distinct wavelength, efficiently couple external lights into the chip. Subsequently, a compact array of 64 microdisk modulators, each with a radius of only 4 µm, encodes the electrical data onto optical carriers via intensity modulation. A low-loss Y-junction–based optical combining tree then merges the 64 modulated carriers into a single bus waveguide. Finally, an inverse-taper edge coupler at the chip output efficiently transfers the spectrally multiplexed signal into a standard single-mode fiber. By admitting 64 spatial inputs while assigning a unique wavelength to each, the architecture simultaneously exploits SDM at the input and WDM at the output, yielding a highly parallelized transmitter with high bandwidth density.

\begin{figure*}[t!]
\centering
\includegraphics[width=0.8\textwidth]{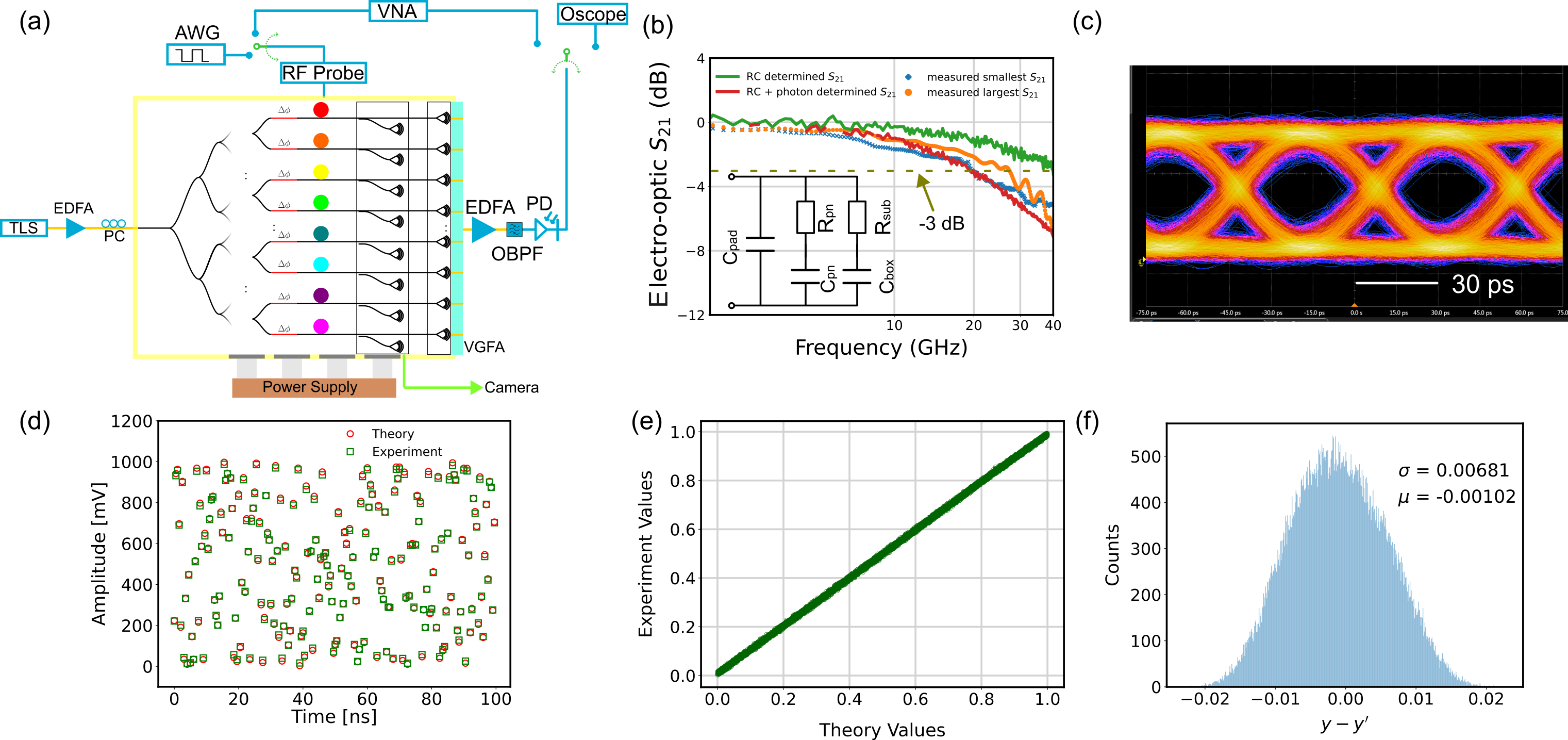}
\caption{High-speed characterization of the Microdisk Transmitter. (a) Schematic of the high-speed measurement setup. OBPF, optical band-pass filter; PD, photodetector; Oscope, oscilloscope; VGFA, V-groove fiber array; VNA, vector network analyzer; AWG, arbitrary waveform generator. (b) Measured (dash, orange and blue) and calculated (line, green and red) normalized electro-optic \(S_{21}\) frequency response. The measured bandwidth of different modulators is between 19 GHz and 28 GHz, which is within the theoretical limitation of the 18 GHz lower band (in resonance wavelength bandwidth) to 42 GHz upper band (RC-only determined bandwidth) bandwidth. Inset: RC circuit model of the modulator that used to fit the measured \(S_{11}\) data. (c) Eye diagram of the transmitter at 20 Gbt/s. (d) Time trace of the measured and expected data encoding results among the transmitter at 20 Gbt/s. (e) Experiment-theory difference standard deviation distribution of the transmitter encoding. The measured standard deviation is around 0.007, indicating an over 7 bits bit precision. (f) Distribution of the theory and experimental encoding results, the theory encoding results are designated as \(y\) and the experimental encoding results are designated as \(y'\).}
\label{fig:3}
\end{figure*}

The parallelized modulator layout overcomes some limitations of the cascaded resonant architectures, such as progressive optical power loss and cumulative thermal and optical crosstalk. By distributing power evenly and arranging modulators in independent optical paths, our design ensures uniform modulation performance across all channels and enables precise, independent tuning and operation. 

\section*{Characterizing the direct current optical response of the Microdisk Transmitter}
To evaluate and analyze the performance of the transmitter, we first performed the measurement on a single-channel microdisk EOIM. Fig.~\ref{fig:2}  shows the measured transmission spectrum of the microdisk EOIM. The average off-resonance IL is 0.6 dB, with a free spectrum range (FSR) of 25.6 nm, a full width at half maximum (FWHM) of 0.29 nm/37 GHz at 0-V d.c. voltage. The apodized TE mode grating coupler has a 3-dB bandwidth of 20 nm. The modulator work in the carrier depletion mode with a reverse biased voltage. Compared with normal lateral junctions, the vertical \textit{p-n} junction microdisk structure enables a low driven voltage by featuring a higher overlap between the electrical  \textit{p–n} depletion region and the microdisk optical whispering gallery mode. The corresponding resonant wavelengths versus the reverse biased voltage are plotted in the Fig.~\ref{fig:2}(d), showing a d.c. response among all these 64 channels with an ER of 9.8-dB, a modulator IL of 2.6-dB, and an OMA of -3.2 dB relative to the input optical power over a 2-V swing voltage.

The microdisk operates in the under-coupled regime, where the external coupling loss is smaller than the round-trip intrinsic loss. Increasing the reverse-biased voltage further depletes the free carriers, thereby reducing free-carrier absorption and lowering the intrinsic loss. This shift pushes the device towards critical coupling, deepening the on-resonance transmission dip and slightly increasing the ER and OMA. Owing to the strong overlap between the optical mode and the \textit{p–n} depletion region, the microdisk exhibits a large electro-optic tuning efficiency of \(90~\mathrm{pm\,V^{-1}}\). The figure of merit for modulation efficiency, \(V_{\pi}\!\cdot\!L\), is therefore \(V_{\pi}\!\cdot\!L = 3.57\,\mathrm{V\!\cdot\!mm}\), which is the best results ever obtained for the carrier depletion mode silicon modulators. By fitting the ring-resonator transmission Lorentz model, the propagation loss is found to decrease from \(\alpha = 8.73~\mathrm{dB\,mm^{-1}}\) to \(\alpha = 6.63~\mathrm{dB\,mm^{-1}}\) as the reverse bias is set from \(0.5~\mathrm{V}\) to \(-1.5~\mathrm{V}\) (Fig.~\ref{fig:2}(f)), and the quality factor \(Q\) rises from 5290 to 6510. We didn't further increase the reverse biased voltage to avoid any potential damages.
The transmission spectrum of the integrated thermal phase shifter, measured with heater voltages from \(0~\mathrm{V}\) to \(5~\mathrm{V}\), is plotted in Fig.~\ref{fig:2}(e). As the drive power increases, the resonance red-shifts. The phase shifter resistance is \(2.0~\mathrm{k}\Omega\), the calculated thermo-optic efficiency and thermal \(\pi\)-phase shift powers are \(420~\mathrm{pm\,mW^{-1}}\) and \(30.5~\mathrm{mW}\), respectively.

\section*{Characterizing the RF response of the Microdisk Transmitter}
The EO bandwidth of the microdisk EOIM is determined by the RC time constant of the junction, the photon lifetime of the microdisk resonator, as well as the offset of the laser wavelength and the resonant wavelength. The RC determined bandwidth is extracted from the measured \(S_{11}\) results. A 40.5 GHz vector network analyzer was used for the S-parameter measurements. The equivalent circuit parameters can be derived through the \(S_{11}\) fitting. In the circuit, \(C_{pn}\) denotes the  \textit{p–n} junction capacitance, \(R_{pn}\) is the  \textit{p–n} junction series resistance, \(C_{box}\) is the buried-oxide capacitance, \(R_{sub}\) stands for the silicon substrate resistance, and \( C_{pad}\) is the capacitance of the pads through the air. The vertical  \textit{p–n} junction has a fitted capacitance \(C_{total}\) of 112.5 fF. Utilizing the equivalent circuits with a \(50\,\Omega\) source impedance, the frequency response across the  \textit{p–n} diode capacitance can be plotted, which corresponds to the small-signal response that is actually effective for the modulation, as shown in Fig.~\ref{fig:3}(b), the RC-time-limited bandwidth of the modulator is \(f_{\mathrm{rc}}\)= 42 GHz. The second EO-bandwidth contributor, the photon lifetime of the microdisk, is characterized by the \(Q\) factor of the resonant cavity. The photon-lifetime-limited bandwidth, \(f_{\mathrm{ph}}\), then can be expressed by $f_{\mathrm{ph}} = \sqrt{\sqrt{2}-1}\,\dfrac{c\,\Delta\lambda}{\lambda^{2}}$~\cite{timurdogan2014ultralow}, where \(c\) is the light speed, and \(\lambda\) is the resonance wavelength. The \(Q\) factor of the microdisk is 5290, so that the \(f_{\mathrm{ph}}\) is around 23 GHz. The final E/O bandwidth of the microdisk can be calculated using \(f_{\mathrm{E/O}} = \dfrac{f_{\mathrm{rc}}\,f_{\mathrm{ph}}}{\sqrt{f_{\mathrm{rc}}^{2}+f_{\mathrm{ph}}^{2}}}\). The fitted \(S_{21}\) is represented by the red and green lines.

The measured electro-optic response of the modulator is shown in Fig.~\ref{fig:3}(b). The laser was aligned off resonance at \(\lambda = 1{,}536.45~\text{nm}\) to preserve linearity and to achieve a faster response by eliminating the photon lifetime influential. A \(50~\text{mV V}_{\mathrm{pp}}\) sinusoidal small signal was applied to the microdisk modulator at frequencies spanning from \(50~\text{MHz}\) to \(40.5~\text{GHz}\). The through-port power was measured with an external high-speed photodetector. The electro-optic 3 dB bandwidth was experimentally measured (orange and blue dash lines) from \(19~\text{GHz}\) to \(28~\text{GHz}\) among different channels. 

\begin{figure*}[t!]
\centering
\includegraphics[width=0.8\textwidth]{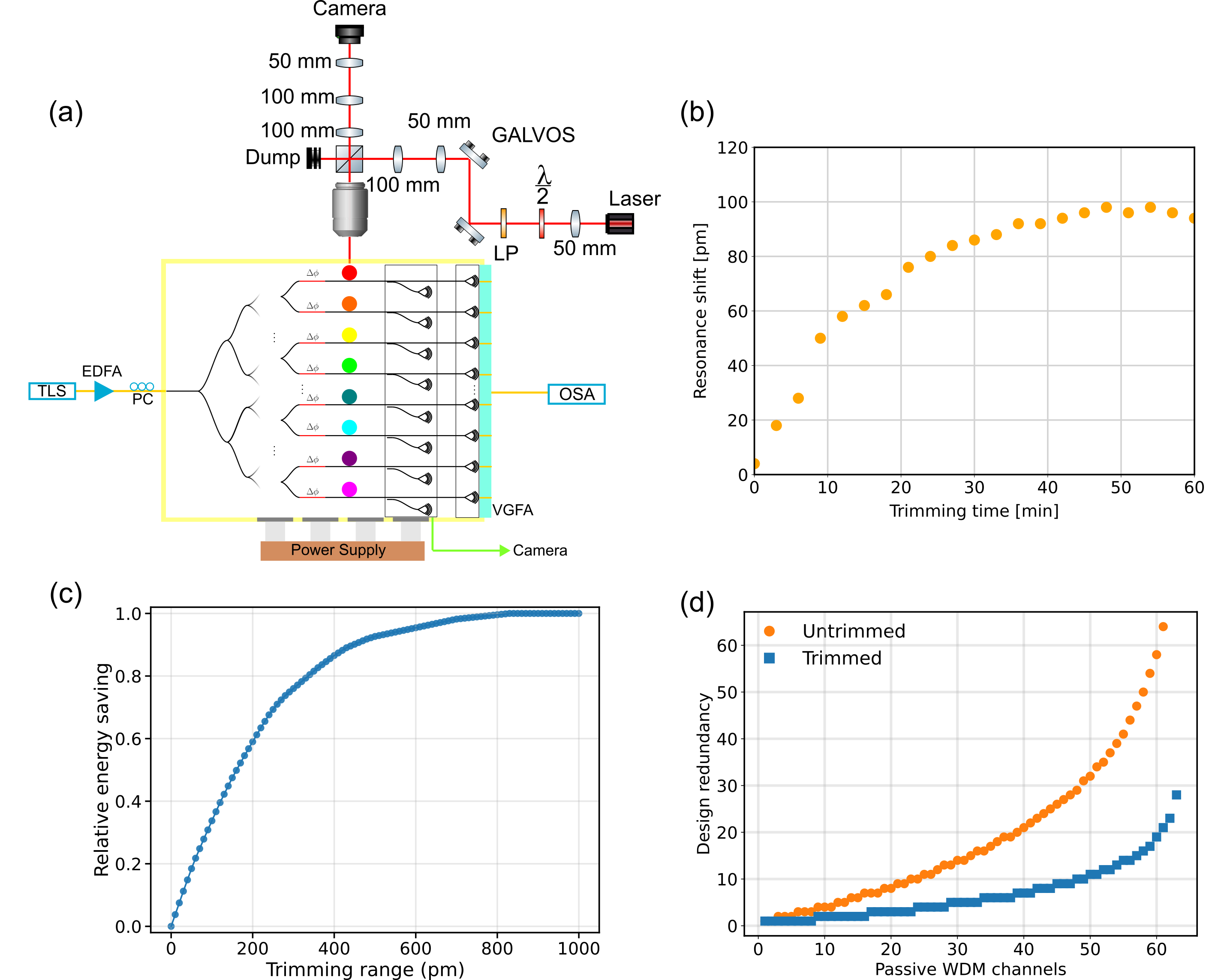}
\caption{Non-volatile trimming of the Microdisk Transmitter. (a) Experimental setup of the non-volatile, close-looped, automatic laser trimming, the 2D grating array was designed to provide in-time monitoring of the emission wavelength. (b) Microdisk resonance wavelength shift versus the trimming time, showing a minimum wavelegnth red shift of 100 pm. (c) Thermal tunning energy saving versus the trimming range. (d) The design redundancy channels needed to achieve the fully passive WDM for pre and post trimmed devices.}
\label{fig:4}
\end{figure*}

\section*{High-speed transmission and energy consumption of the Microdisk Transmitter}
High-speed optical eye diagrams, presented in Fig.~\ref{fig:3}(c), were obtained utilizing a high-speed electro–optic measurement setup with a real-time oscilloscope at data rates of 20 ~Gbt/s (setup-limited). The open eye diagram indicates a good signal integrity and robust modulation performance. The 64-channel photonic transmitter demonstrates an high bandwidth of 1.28 ~Tbt/s while maintaining a low dynamic energy consumption of approximately \(29~\text{fJ/bit}\) per channel, calculated using the conventional \(CV^{2}/4\) metric with a 1-V swing voltage. It is worth noting that 20~Gbt/s eyes are the upper limit of the experimental setup, a higher-bandwidth setup can further improve the transmitter capacity \cite{daudlin2025three}.

To quantitatively evaluate the modulation accuracy, we performed comprehensive encoding and decoding characterization of the modulator. Data signals generated by the AWG were first pulse-amplitude-modulated to $2~\mathrm{Gbt/s}$ and then encoded onto the EOIM, transmitted through optical fiber, amplified by an EDFA, detected by a high-speed photodetector, and digitized using a high-speed oscilloscope. The modulation accuracy was quantitatively assessed by calculating the standard deviation of the difference between the theoretically generated AWG data and the measured oscilloscope waveforms. The measured standard deviation of the difference was $0.007$, corresponding to an effective bit precision of $7~\text{bits}$. This low standard deviation highlights the transmitter's high modulation accuracy and stability, confirming its suitability for high-performance optical communication and computing systems.

\section*{Non-volatile trimming of the Microdisk Transmitter}

\begin{figure*}[t!]
\centering
\includegraphics[width=0.9\textwidth]{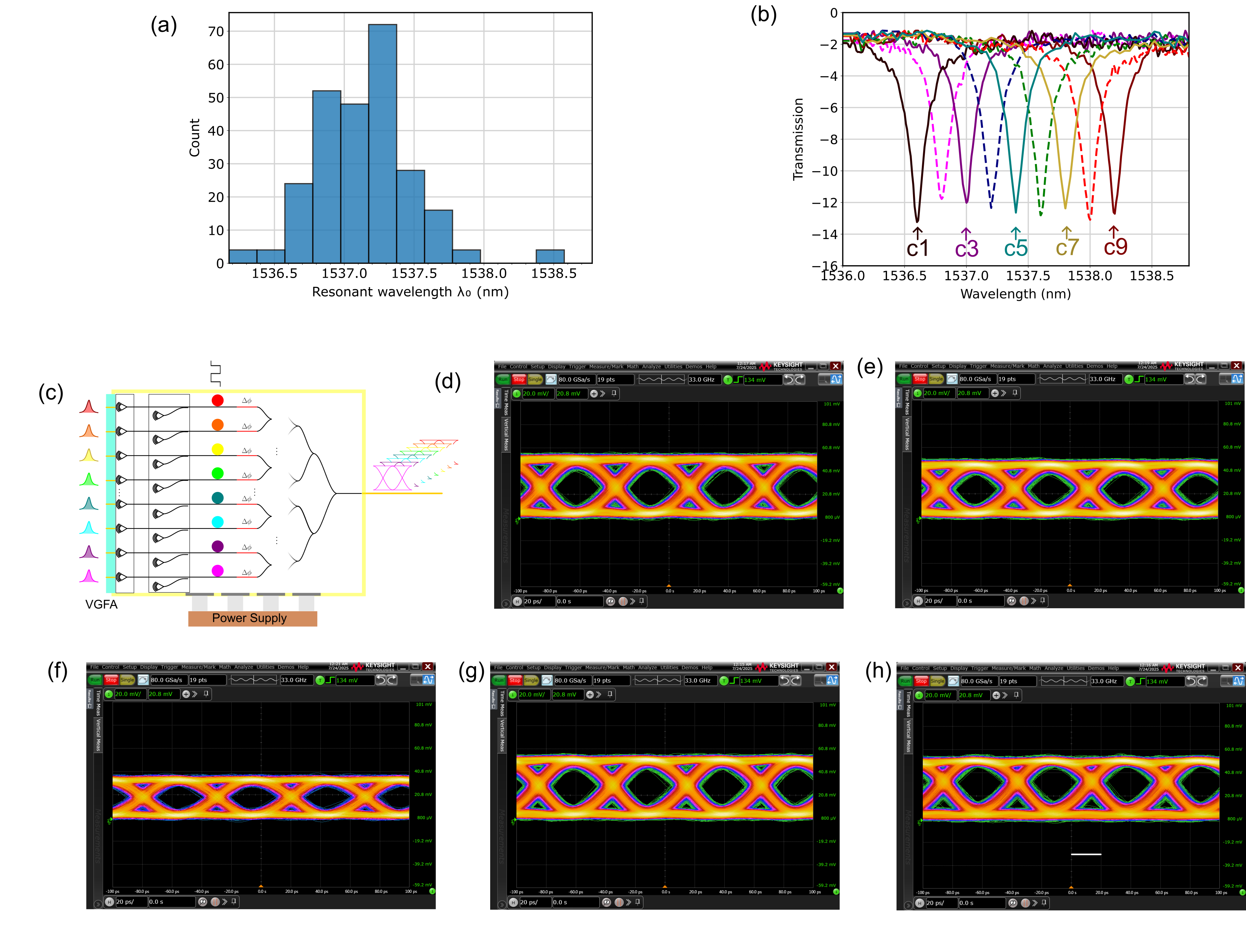}
\caption{ Fully passive DWDM transmission. (a) Wavelength distribution of the foundry fabricated microdisk chip. (b) Selected DWDM wavelength distribution after the trimming, showing uniform wavelength spacing of 25 GHz. (c-h) Schematic and results of the 20 Gbt/s eye-diagram of the trimmed microdisk modulators, which demonstrated a fully passive, dense wavelength division (50 GHz) channel, scalar bar: 40 ps.}
\label{fig:5}
\end{figure*}

The microdisk transmitter chip is designed to support both high parallelism SDM and high bandwidth density WDM. WDM requires each microdisk modulating at a distinct dedicated wavelength. On the other hand, fabrication-induced variations in resonance frequency is a major issue in integrated photonic resonant devices and this tolerance cannot be solely achieved through optimized fabrication. While traditional thermal tuning methods are used to compensate for these variations, they suffer from high power consumption. To retune the system to dedicated target resonance, we developed a nonvolatile resonance tunning setup based on the laser trimming. Specifically, our trimming method worked for ultra-compact microdisk with top SiO\(_2\) cladding and metal embedding by changing the top cladding compaction, enabling automated in-situ, close-looped, non-reversal, low-loss, and picometer-precision permanent wavelength tuning, significantly enhancing the controllability compared to previous methods.

To accomplish the laser-based permanent shifting of microdisk resonances, we introduce an additional free space path to controllably inject a trimming laser onto each ring. This path includes a green laser and critically, a two axis galvanometers for controllable beam focusing and routing (Thorlabs GSV202). This path is merged into the existing chip characterization paths by using an angled half reflective mirror before the objective. We discretize this path into 360 steps around the circle. Every iteration step along the circle takes \(\sim 1~\mathrm{s}\), the trimmed spectrum was recorded after each half circle, and summarized in Fig.~\ref{fig:4}b. Due to the strong mode confinement of vertical silicon p-n junction, current trimming mechanism has a small wavelength tuning range from 100 pm to 380 pm.

While the current wavelength trimming range is limited, this trimming technique offers two significant practical applications. The first application is reducing the overall energy consumption in WDM transmission systems to maintain the resonant wavelength. By employing the current trimming method, the energy consumption required for tuning and maintaining the target wavelength can be effectively reduced by 33 \%, as illustrated in Fig.~\ref{fig:4}c.

Another important application is reducing the required design redundancy of the microdisk resonators. In temperature-sensitive applications, such as label-free biosensing (e.g., microring sensors in liquid) \cite{butt2025integrated, kazanskiy2023review} and cryogenic quantum platforms \cite{chakraborty2020cryogenic}, the active thermal tuning will influence the experimental results, multiple redundant resonators are typically fabricated for a target wavelength to statistically ensure at least one ring naturally aligns to the desired resonance. We set the design redundancy as the cost function, and the laser trimming was employed to reduce it. By employing laser trimming, the necessary redundancy was reduced, effectively halving cost function, as illustrated in Fig.~\ref{fig:4}d.

\section*{Fully passive DWDM using the Microdisk Transmitter}

Fig.~\ref{fig:5}a shows the wavelength distribution of the foundry fabricated microdisk modulators, these modulators are designed to have the same target resonant wavelength at 1537.2 nm. The actual measured transmission spectrum is separated from 1536.5 nm to 1539 nm. Combined with the non-volatile trimming, this initial distribution can be employed to build the high bandwidth WDM transmission link. We select the channel with resonances close to the DWDM ITU wavelength, and use the trimming setup to perform automated precisely laser annealing, after trimming, these channels are evenly distributed (\(25~\text{GHz}\) spacing) within one FSR. 

The OMA, bandwidth and eye-diagram of the laser-trimmed DWDM modulators were experimentally characterized, revealing no discernible degradation after trimming. To mitigate inter-channel crosstalk, channels 1, 3, 5, 7, and 9 were specifically selected for DWDM transmission experiments. The implemented DWDM system features a standard wavelength channel spacing of 50 GHz, with each selected channel independently modulated at a data rate of $20~\mathrm{Gbt/s}$. Consequently, the chip support a fully passive, 5-channel DWDM system with per-fiber capacity of $100~\mathrm{Gbt/s}$. These results underscore the robust performance and scalability of the proposed trimmed EOIM-based transmitter, highlighting its potential as a key component in high-capacity integrated photonic communication networks.

\section*{Discussion and outlook }
Integrated SiPh chips provide a promising platform to meet the escalating demands for high-speed and energy-efficient data movement in modern data centers, particularly for AI applications. Here, we realize this promise by demonstrating a large-scale SiPh transmitter comprising an array of 64 EOIMs. Fabricated on a 300-mm wafer via a fully CMOS-compatible process, each EOIM, based on a compact microdisk resonator with an 8-µm diameter, achieves a high electro-optic tuning efficiency of $90~\mathrm{pm\,V^{-1}}$, a 3-dB electro-optic bandwidth of $20~\mathrm{GHz}$, an ER of $9.8~\mathrm{dB}$, an IL of $2.6~\mathrm{dB}$, and an OMA of -3.2 dB relative to input optical power at a $2~\mathrm{V}$ swing voltage. Moreover, the microdisk modulator exhibits a low $V_{\pi}\cdot L$ of $3.57~\mathrm{V\cdot mm}$. These characteristics enable high-precision, energy-efficient dynamic modulation at $29~\mathrm{fJ/bit}$, supporting data rates up to $20~\mathrm{Gbt/s}$ per channel.

The demonstrated transmitter achieves a data transmission capacity of $1.28~\mathrm{Tbt/s}$, which was limited by our characterization equipment. By utilizing advanced, high-speed AWG sources, modulation speeds of $30~\mathrm{Gbt/s}$ per channel are achievable, enabling scalability to a capacity of $1.8~\mathrm{Tbt/s}$ \cite{daudlin2025three}.

Furthermore, we demonstrated that fabrication-induced resonance-frequency variations in the microdisk modulators can be mitigated through an in-situ automated, clos-looped, permanent laser-trimming technique with picometer-scale precision. Utilizing our customized, nonvolatile laser-trimming platform, a passive 5-channel DWDM transmitter system is demonstrated. This trimming method introduces negligible optical losses and preserves the high modulation performance. The current trimming technology reduces 33 \% thermal energy consumption needed to tune the foundry microdisk to target resonant wavelength, and effectively halving the number of redundant microdisks needed for temperature sensitive applications where thermal tunning is not allowed. When combined with emerging integrated chip-scale optical frequency comb sources, our scalable DWDM architecture paves the way toward ultra-high-density silicon photonic interconnects capable of supporting massively parallel optical communications.

\section*{methods}

\subsection*{Device packaging}
\label{sec:m2}
Electrical packaging was performed by Optelligent. The active silicon photonic chip was first die-bonded to a co-designed printed circuit board (PCB). The PCB bond pads were finished with PCBWay. All d.c. and thermal pads around the periphery of the photonic chips were wirebonded to the PCB and the internal RF pads were left exposed for high speed probing. No optical packaging was performed for the current chip, the input fiber is standard single mode fiber, the output 1D grating was recorded by fiber, and the 2D grating output was parallel recorded by an IR camera.

\subsection*{Direct current characterization of the transmitter}
\label{sec:m3}
The transmitter d.c. electro-optic response was performed in the packaged photonic chip. A tunable Ando laser 4321D and an Ando OSA 6317 are synchronized with each other to record the transmission spectrum. By apply a voltage to the PCB board to reverse bias the modulator at varying d.c. voltages and record the response spectrum under different voltages, we can extract the d.c. response of the modulator at different voltages over a wide spectrum range.

\subsection*{Bandwidth measurement of the transmitter}
\label{sec:m4}
A tunable CW laser, Ando 4321D, was pre-amplified using EDFA and then send to the polarization controller, the light was then coupled to the chip using edge coupler. The laser wavelength is located at the power transfer function quadrature point. For small-signal bandwidth characterization, the calibrated VNA supplied a swept-frequency stimulus from 50 MHz to 40.5 GHz, the VNA signal was delivered to the modulator via a ground–signal(GS) microwave probe. The modulated optical signal was detected by a 50 GHz photodetector, and then send to the receiver channel of the VNA, enabling measurement of the electro-optic bandwidth. The frequency response of the photodetector was obtained from its datasheet, and it was used to compensate for the measured S\(_2\)\(_1\) data to achieve the frequency response of the microdisk modulator only.

\subsection*{High-speed eye-diagram measurement and bit precision measurement of the transmitter}
\label{sec:m5}
A tunable CW laser, Ando 4321D, was pre-amplified using EDFA and then send to the polarization controller, the light was then coupled to the chip using edge coupler. The laser wavelength is located at the transfer function quadrature point. The 20-Gbt/s electrical signal that drives the modulator was generated by program the AWG with an pseudo random binary sequence (PRBS). The highest swing voltage Vpp of the AWG is 0.5 V, therefore a 20 GHz electrical power amplifiers were added to amplify the swing voltage to 1-V. The RF drive signal were applied to the transmitter using probe station. The microdisk output light was amplified by another EDFA to compensate for the link losses, a narrow band 100 GHz filter was used to remove the ASE noise. The filtered signal was then received by the high speed photodetector and oscilloscope for the eye diagram measurement. The clock and trigger to the oscilloscope came from another output channel of the AWG.

This setup was also used for the bit precision measurement. The AWG drive signal was changed from PRBS to random analog values between 0 and 1, all these signal have the same time period, which is determined by the AWG sampling rates. The data were extracted from the oscilloscope, and compared with the original ground-truth AWG drive signal. The bit precision was calculated by calculating the mutual information of the original AWG signal and the extracted oscilloscope data.

\subsection*{Laser trimming Setup}
\label{sec:m6}
Tuning of the trimming spot is accomplished by scanning the galvanometers angles and makes the first-order beam touted across the waveguide in a dithered circular pattern. We discretize this path into 360 steps around the circle. Every iteration step along the circle takes \(\sim 1~\mathrm{s}\). Once the laser intensity and the galvanometers scanning parameters are set, we leave the trimming light constantly on.

\begin{figure}[!htbp]
  \centering
  \includegraphics[width=0.4\textwidth]{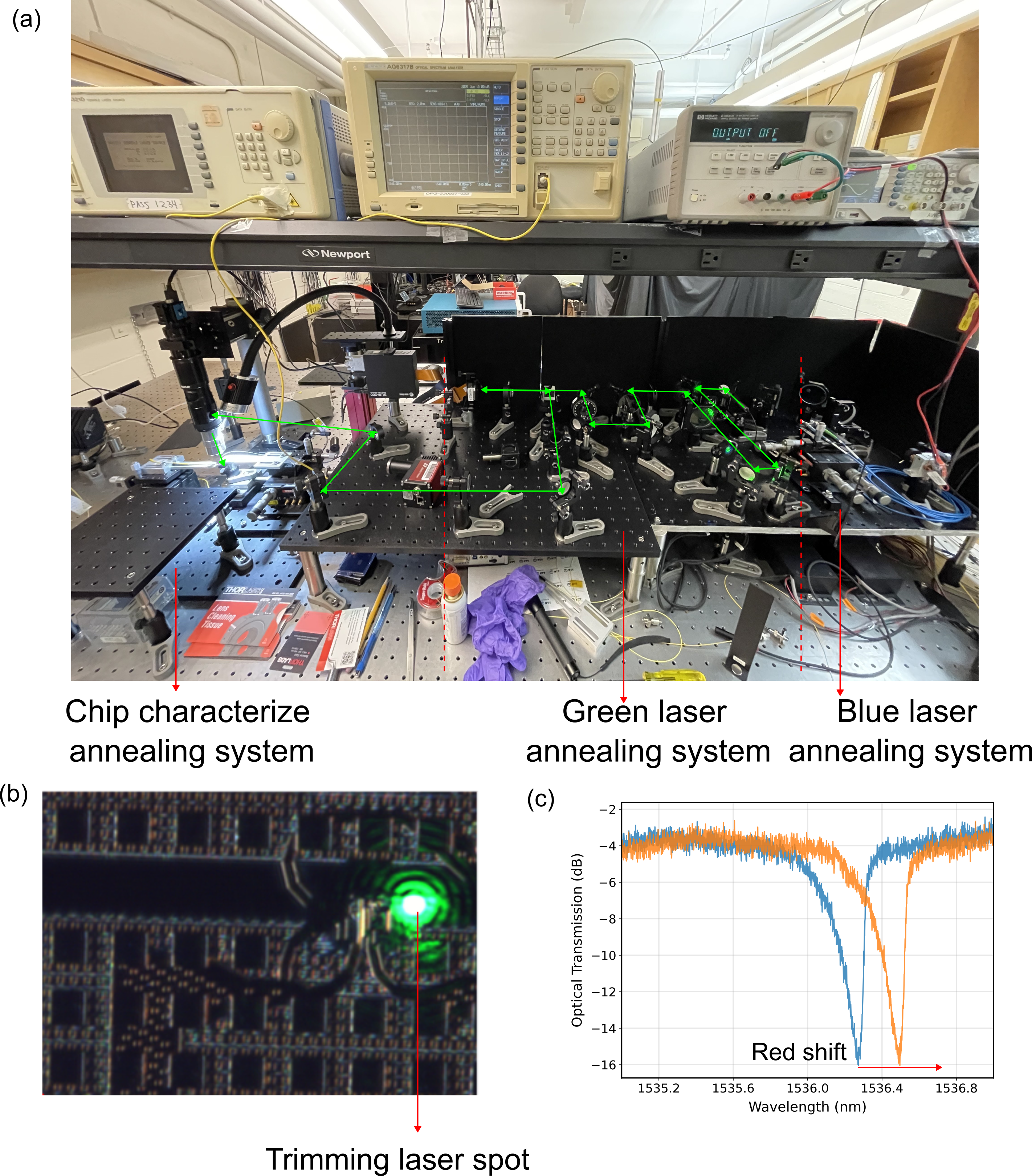}
  \caption{Laser annealing setup and results for the Silicon microdisk E/O modulator, in the experimental setup, we have replaced the galvanometers to spatial light modulator, which support multiple, in parallel trimming, the technique is under development.}
  \label{fig:sfig42}
\end{figure}

After the trimming, wavelength red-shifted is observed, however, due to the strong mode confinement of the silicon waveguide, the laser trimming induced SiO\(_2\)\ compaction change has low influential for the mode effective index change. So, to target the devices for a specific wavelength we select a chip with resonances close to the desired wavelength. This minimizes the required trimming power/duration. In addition, we observe no significant change in the extracted full width at half maximum given the data in Figure~\ref{fig:sfig42}. We measure the laser-trimmed devices again after roughly two months of ambient storage and observe no relaxation in trimming. No damage was observed in the microdisk, the E/O modulation and thermal phase shifting still operate well after the laser trimming. 

It needed to be point out that the current trimming technology is achieved through changing the SiO\(_2\)\ cladding compaction, the trimming efficiency is related to the SiO\(_2\)\ cladding quality. Using the same trimming methods, we achieve over 0.35 nm wavelength red tunning on some device, and the 0.1 nm wavelength red tuning we used in the manuscript is the minimum wavelength red tunning we obtained among different devices.

\section*{Acknowledgements}

The authors gratefully acknowledge the slmsuite team for their contributions to the trimming algorithm and code development.

\subsection*{Funding}
This work was funded by the research collaboration agreements with DARPA Nanowatt platform for Sensing, Analysis, and Computation (NaPSAC) platform, Nippon Telegraph and Telephone (NTT) research, and Taiwan Semiconductor Manufacturing Company (TSMC).

\subsection*{Contributions}
Chao Luan built the experiment setup for the chip characterization, high-speed transmission link, and laser trimming, wrote the code and algorithm (with help from the slmsuite), conducted the experiment, performed the data analysis, and wrote the manuscript. Alex Sludds designed the chip, and performed initial device testing. Ian Christen assisted with the chip package and laser trimming setup. Chao Li developed the non-volatile trimming method. Ryan Hamerly and Dirk Englund supervised the project. Ryan Hamerly and Dirk Englund revised the manuscript. 

\section*{Competing Interests}
The authors Chao Luan, Ryan Hamerly, and Dirk Englund  disclose that they are inventors on pending patent US Application No. 63/768502 where MIT is the patent applicant, which covers the Demonstration and Non-volatile trimming of the transmitter chip described in this work.

\section*{Data and Materials Availability}
The data from this work will be made available upon reasonable request.

\section*{Code Availability}
The code used to characterize the chip and perform laser trimming from this work will be made available upon reasonable request, with public portions available at \url{github.com/slmsuite/slmsuite}.

\bibliography{main} 

\begin{thebibliography}{61}%
\makeatletter
\providecommand \@ifxundefined [1]{%
 \@ifx{#1\undefined}
}%
\providecommand \@ifnum [1]{%
 \ifnum #1\expandafter \@firstoftwo
 \else \expandafter \@secondoftwo
 \fi
}%
\providecommand \@ifx [1]{%
 \ifx #1\expandafter \@firstoftwo
 \else \expandafter \@secondoftwo
 \fi
}%
\providecommand \natexlab [1]{#1}%
\providecommand \enquote  [1]{``#1''}%
\providecommand \bibnamefont  [1]{#1}%
\providecommand \bibfnamefont [1]{#1}%
\providecommand \citenamefont [1]{#1}%
\providecommand \href@noop [0]{\@secondoftwo}%
\providecommand \href [0]{\begingroup \@sanitize@url \@href}%
\providecommand \@href[1]{\@@startlink{#1}\@@href}%
\providecommand \@@href[1]{\endgroup#1\@@endlink}%
\providecommand \@sanitize@url [0]{\catcode `\\12\catcode `\$12\catcode `\&12\catcode `\#12\catcode `\^12\catcode `\_12\catcode `\%12\relax}%
\providecommand \@@startlink[1]{}%
\providecommand \@@endlink[0]{}%
\providecommand \url  [0]{\begingroup\@sanitize@url \@url }%
\providecommand \@url [1]{\endgroup\@href {#1}{\urlprefix }}%
\providecommand \urlprefix  [0]{URL }%
\providecommand \Eprint [0]{\href }%
\providecommand \doibase [0]{https://doi.org/}%
\providecommand \selectlanguage [0]{\@gobble}%
\providecommand \bibinfo  [0]{\@secondoftwo}%
\providecommand \bibfield  [0]{\@secondoftwo}%
\providecommand \translation [1]{[#1]}%
\providecommand \BibitemOpen [0]{}%
\providecommand \bibitemStop [0]{}%
\providecommand \bibitemNoStop [0]{.\EOS\space}%
\providecommand \EOS [0]{\spacefactor3000\relax}%
\providecommand \BibitemShut  [1]{\csname bibitem#1\endcsname}%
\let\auto@bib@innerbib\@empty
\bibitem [{\citenamefont {Miller}(2000)}]{867687}%
  \BibitemOpen
  \bibfield  {author} {\bibinfo {author} {\bibfnamefont {D.}~\bibnamefont {Miller}},\ }\bibfield  {title} {\bibinfo {title} {Rationale and challenges for optical interconnects to electronic chips},\ }\href {https://doi.org/10.1109/5.867687} {\bibfield  {journal} {\bibinfo  {journal} {Proceedings of the IEEE}\ }\textbf {\bibinfo {volume} {88}},\ \bibinfo {pages} {728} (\bibinfo {year} {2000})}\BibitemShut {NoStop}%
\bibitem [{\citenamefont {Miller}(2017)}]{7805240}%
  \BibitemOpen
  \bibfield  {author} {\bibinfo {author} {\bibfnamefont {D.~A.~B.}\ \bibnamefont {Miller}},\ }\bibfield  {title} {\bibinfo {title} {Attojoule optoelectronics for low-energy information processing and communications},\ }\href {https://doi.org/10.1109/JLT.2017.2647779} {\bibfield  {journal} {\bibinfo  {journal} {Journal of Lightwave Technology}\ }\textbf {\bibinfo {volume} {35}},\ \bibinfo {pages} {346} (\bibinfo {year} {2017})}\BibitemShut {NoStop}%
\bibitem [{\citenamefont {Cheng}\ \emph {et~al.}(2018)\citenamefont {Cheng}, \citenamefont {Bahadori}, \citenamefont {Glick}, \citenamefont {Rumley},\ and\ \citenamefont {Bergman}}]{cheng2018recent}%
  \BibitemOpen
  \bibfield  {author} {\bibinfo {author} {\bibfnamefont {Q.}~\bibnamefont {Cheng}}, \bibinfo {author} {\bibfnamefont {M.}~\bibnamefont {Bahadori}}, \bibinfo {author} {\bibfnamefont {M.}~\bibnamefont {Glick}}, \bibinfo {author} {\bibfnamefont {S.}~\bibnamefont {Rumley}},\ and\ \bibinfo {author} {\bibfnamefont {K.}~\bibnamefont {Bergman}},\ }\bibfield  {title} {\bibinfo {title} {Recent advances in optical technologies for data centers: a review},\ }\href@noop {} {\bibfield  {journal} {\bibinfo  {journal} {Optica}\ }\textbf {\bibinfo {volume} {5}},\ \bibinfo {pages} {1354} (\bibinfo {year} {2018})}\BibitemShut {NoStop}%
\bibitem [{\citenamefont {Wang}\ \emph {et~al.}(2023)\citenamefont {Wang}, \citenamefont {Qin}, \citenamefont {Jacobs}, \citenamefont {Holmes}, \citenamefont {Rajbhandari}, \citenamefont {Ruwase}, \citenamefont {Yan}, \citenamefont {Yang},\ and\ \citenamefont {He}}]{wang2023extremely}%
  \BibitemOpen
  \bibfield  {author} {\bibinfo {author} {\bibfnamefont {G.}~\bibnamefont {Wang}}, \bibinfo {author} {\bibfnamefont {H.}~\bibnamefont {Qin}}, \bibinfo {author} {\bibfnamefont {S.}~\bibnamefont {Jacobs}}, \bibinfo {author} {\bibfnamefont {C.}~\bibnamefont {Holmes}}, \bibinfo {author} {\bibfnamefont {S.}~\bibnamefont {Rajbhandari}}, \bibinfo {author} {\bibfnamefont {O.}~\bibnamefont {Ruwase}}, \bibinfo {author} {\bibfnamefont {F.}~\bibnamefont {Yan}}, \bibinfo {author} {\bibfnamefont {L.}~\bibnamefont {Yang}},\ and\ \bibinfo {author} {\bibfnamefont {Y.~Z.}\ \bibnamefont {He}},\ }\bibfield  {title} {\bibinfo {title} {Extremely efficient collective communication for giant model training},\ }\href@noop {} {\bibfield  {journal} {\bibinfo  {journal} {arXiv preprint arXiv:2306.10209}\ } (\bibinfo {year} {2023})}\BibitemShut {NoStop}%
\bibitem [{\citenamefont {Van~Heddeghem}\ \emph {et~al.}(2014)\citenamefont {Van~Heddeghem}, \citenamefont {Lambert}, \citenamefont {Lannoo}, \citenamefont {Colle}, \citenamefont {Pickavet},\ and\ \citenamefont {Demeester}}]{van2014trends}%
  \BibitemOpen
  \bibfield  {author} {\bibinfo {author} {\bibfnamefont {W.}~\bibnamefont {Van~Heddeghem}}, \bibinfo {author} {\bibfnamefont {S.}~\bibnamefont {Lambert}}, \bibinfo {author} {\bibfnamefont {B.}~\bibnamefont {Lannoo}}, \bibinfo {author} {\bibfnamefont {D.}~\bibnamefont {Colle}}, \bibinfo {author} {\bibfnamefont {M.}~\bibnamefont {Pickavet}},\ and\ \bibinfo {author} {\bibfnamefont {P.}~\bibnamefont {Demeester}},\ }\bibfield  {title} {\bibinfo {title} {Trends in worldwide ict electricity consumption from 2007 to 2012},\ }\href@noop {} {\bibfield  {journal} {\bibinfo  {journal} {Computer communications}\ }\textbf {\bibinfo {volume} {50}},\ \bibinfo {pages} {64} (\bibinfo {year} {2014})}\BibitemShut {NoStop}%
\bibitem [{\citenamefont {Frank}\ \emph {et~al.}(2006)\citenamefont {Frank}, \citenamefont {Haensch}, \citenamefont {Shahidi},\ and\ \citenamefont {Dokumaci}}]{frank2006optimizing}%
  \BibitemOpen
  \bibfield  {author} {\bibinfo {author} {\bibfnamefont {D.~J.}\ \bibnamefont {Frank}}, \bibinfo {author} {\bibfnamefont {W.}~\bibnamefont {Haensch}}, \bibinfo {author} {\bibfnamefont {G.}~\bibnamefont {Shahidi}},\ and\ \bibinfo {author} {\bibfnamefont {O.~H.}\ \bibnamefont {Dokumaci}},\ }\bibfield  {title} {\bibinfo {title} {Optimizing cmos technology for maximum performance},\ }\href@noop {} {\bibfield  {journal} {\bibinfo  {journal} {IBM journal of research and development}\ }\textbf {\bibinfo {volume} {50}},\ \bibinfo {pages} {419} (\bibinfo {year} {2006})}\BibitemShut {NoStop}%
\bibitem [{\citenamefont {Waldrop}(2016)}]{waldrop2016more}%
  \BibitemOpen
  \bibfield  {author} {\bibinfo {author} {\bibfnamefont {M.~M.}\ \bibnamefont {Waldrop}},\ }\bibfield  {title} {\bibinfo {title} {More than moore},\ }\href@noop {} {\bibfield  {journal} {\bibinfo  {journal} {Nature}\ }\textbf {\bibinfo {volume} {530}},\ \bibinfo {pages} {144} (\bibinfo {year} {2016})}\BibitemShut {NoStop}%
\bibitem [{\citenamefont {Keckler}\ \emph {et~al.}(2011)\citenamefont {Keckler}, \citenamefont {Dally}, \citenamefont {Khailany}, \citenamefont {Garland},\ and\ \citenamefont {Glasco}}]{keckler2011gpus}%
  \BibitemOpen
  \bibfield  {author} {\bibinfo {author} {\bibfnamefont {S.~W.}\ \bibnamefont {Keckler}}, \bibinfo {author} {\bibfnamefont {W.~J.}\ \bibnamefont {Dally}}, \bibinfo {author} {\bibfnamefont {B.}~\bibnamefont {Khailany}}, \bibinfo {author} {\bibfnamefont {M.}~\bibnamefont {Garland}},\ and\ \bibinfo {author} {\bibfnamefont {D.}~\bibnamefont {Glasco}},\ }\bibfield  {title} {\bibinfo {title} {Gpus and the future of parallel computing},\ }\href@noop {} {\bibfield  {journal} {\bibinfo  {journal} {IEEE micro}\ }\textbf {\bibinfo {volume} {31}},\ \bibinfo {pages} {7} (\bibinfo {year} {2011})}\BibitemShut {NoStop}%
\bibitem [{\citenamefont {Lee}\ \emph {et~al.}(2022)\citenamefont {Lee}, \citenamefont {Nedovic}, \citenamefont {Greer},\ and\ \citenamefont {Gray}}]{lee2022beyond}%
  \BibitemOpen
  \bibfield  {author} {\bibinfo {author} {\bibfnamefont {B.~G.}\ \bibnamefont {Lee}}, \bibinfo {author} {\bibfnamefont {N.}~\bibnamefont {Nedovic}}, \bibinfo {author} {\bibfnamefont {T.~H.}\ \bibnamefont {Greer}},\ and\ \bibinfo {author} {\bibfnamefont {C.~T.}\ \bibnamefont {Gray}},\ }\bibfield  {title} {\bibinfo {title} {Beyond cpo: A motivation and approach for bringing optics onto the silicon interposer},\ }\href@noop {} {\bibfield  {journal} {\bibinfo  {journal} {Journal of Lightwave Technology}\ }\textbf {\bibinfo {volume} {41}},\ \bibinfo {pages} {1152} (\bibinfo {year} {2022})}\BibitemShut {NoStop}%
\bibitem [{\citenamefont {Markov}(2014)}]{markov2014limits}%
  \BibitemOpen
  \bibfield  {author} {\bibinfo {author} {\bibfnamefont {I.~L.}\ \bibnamefont {Markov}},\ }\bibfield  {title} {\bibinfo {title} {Limits on fundamental limits to computation},\ }\href@noop {} {\bibfield  {journal} {\bibinfo  {journal} {Nature}\ }\textbf {\bibinfo {volume} {512}},\ \bibinfo {pages} {147} (\bibinfo {year} {2014})}\BibitemShut {NoStop}%
\bibitem [{\citenamefont {Baliga}\ \emph {et~al.}(2011)\citenamefont {Baliga}, \citenamefont {Ayre}, \citenamefont {Hinton},\ and\ \citenamefont {Tucker}}]{baliga2011energy}%
  \BibitemOpen
  \bibfield  {author} {\bibinfo {author} {\bibfnamefont {J.}~\bibnamefont {Baliga}}, \bibinfo {author} {\bibfnamefont {R.}~\bibnamefont {Ayre}}, \bibinfo {author} {\bibfnamefont {K.}~\bibnamefont {Hinton}},\ and\ \bibinfo {author} {\bibfnamefont {R.~S.}\ \bibnamefont {Tucker}},\ }\bibfield  {title} {\bibinfo {title} {Energy consumption in wired and wireless access networks},\ }\href@noop {} {\bibfield  {journal} {\bibinfo  {journal} {IEEE Communications Magazine}\ }\textbf {\bibinfo {volume} {49}},\ \bibinfo {pages} {70} (\bibinfo {year} {2011})}\BibitemShut {NoStop}%
\bibitem [{\citenamefont {Association}\ \emph {et~al.}(2005)\citenamefont {Association} \emph {et~al.}}]{semiconductor2005international}%
  \BibitemOpen
  \bibfield  {author} {\bibinfo {author} {\bibfnamefont {S.~I.}\ \bibnamefont {Association}} \emph {et~al.},\ }\bibfield  {title} {\bibinfo {title} {International technology roadmap for semiconductors 2005 edition},\ }\href@noop {} {\bibfield  {journal} {\bibinfo  {journal} {http://www. itrs. net/Links/2005ITRS/ExecSum2005. pdf}\ } (\bibinfo {year} {2005})}\BibitemShut {NoStop}%
\bibitem [{\citenamefont {Hochberg}\ and\ \citenamefont {Baehr-Jones}(2010)}]{hochberg2010towards}%
  \BibitemOpen
  \bibfield  {author} {\bibinfo {author} {\bibfnamefont {M.}~\bibnamefont {Hochberg}}\ and\ \bibinfo {author} {\bibfnamefont {T.}~\bibnamefont {Baehr-Jones}},\ }\bibfield  {title} {\bibinfo {title} {Towards fabless silicon photonics},\ }\href@noop {} {\bibfield  {journal} {\bibinfo  {journal} {Nature photonics}\ }\textbf {\bibinfo {volume} {4}},\ \bibinfo {pages} {492} (\bibinfo {year} {2010})}\BibitemShut {NoStop}%
\bibitem [{\citenamefont {Tucker}\ \emph {et~al.}(2009)\citenamefont {Tucker}, \citenamefont {Parthiban}, \citenamefont {Baliga}, \citenamefont {Hinton}, \citenamefont {Ayre},\ and\ \citenamefont {Sorin}}]{tucker2009evolution}%
  \BibitemOpen
  \bibfield  {author} {\bibinfo {author} {\bibfnamefont {R.~S.}\ \bibnamefont {Tucker}}, \bibinfo {author} {\bibfnamefont {R.}~\bibnamefont {Parthiban}}, \bibinfo {author} {\bibfnamefont {J.}~\bibnamefont {Baliga}}, \bibinfo {author} {\bibfnamefont {K.}~\bibnamefont {Hinton}}, \bibinfo {author} {\bibfnamefont {R.~W.}\ \bibnamefont {Ayre}},\ and\ \bibinfo {author} {\bibfnamefont {W.~V.}\ \bibnamefont {Sorin}},\ }\bibfield  {title} {\bibinfo {title} {Evolution of wdm optical ip networks: A cost and energy perspective},\ }\href@noop {} {\bibfield  {journal} {\bibinfo  {journal} {Journal of Lightwave Technology}\ }\textbf {\bibinfo {volume} {27}},\ \bibinfo {pages} {243} (\bibinfo {year} {2009})}\BibitemShut {NoStop}%
\bibitem [{\citenamefont {Agrell}\ \emph {et~al.}(2016)\citenamefont {Agrell}, \citenamefont {Karlsson}, \citenamefont {Chraplyvy}, \citenamefont {Richardson}, \citenamefont {Krummrich}, \citenamefont {Winzer}, \citenamefont {Roberts}, \citenamefont {Fischer}, \citenamefont {Savory}, \citenamefont {Eggleton} \emph {et~al.}}]{agrell2016roadmap}%
  \BibitemOpen
  \bibfield  {author} {\bibinfo {author} {\bibfnamefont {E.}~\bibnamefont {Agrell}}, \bibinfo {author} {\bibfnamefont {M.}~\bibnamefont {Karlsson}}, \bibinfo {author} {\bibfnamefont {A.}~\bibnamefont {Chraplyvy}}, \bibinfo {author} {\bibfnamefont {D.~J.}\ \bibnamefont {Richardson}}, \bibinfo {author} {\bibfnamefont {P.~M.}\ \bibnamefont {Krummrich}}, \bibinfo {author} {\bibfnamefont {P.}~\bibnamefont {Winzer}}, \bibinfo {author} {\bibfnamefont {K.}~\bibnamefont {Roberts}}, \bibinfo {author} {\bibfnamefont {J.~K.}\ \bibnamefont {Fischer}}, \bibinfo {author} {\bibfnamefont {S.~J.}\ \bibnamefont {Savory}}, \bibinfo {author} {\bibfnamefont {B.~J.}\ \bibnamefont {Eggleton}}, \emph {et~al.},\ }\bibfield  {title} {\bibinfo {title} {Roadmap of optical communications},\ }\href@noop {} {\bibfield  {journal} {\bibinfo  {journal} {Journal of optics}\ }\textbf {\bibinfo {volume} {18}},\ \bibinfo {pages} {063002} (\bibinfo {year} {2016})}\BibitemShut {NoStop}%
\bibitem [{\citenamefont {Rizzo}\ \emph {et~al.}(2023)\citenamefont {Rizzo}, \citenamefont {Novick}, \citenamefont {Gopal}, \citenamefont {Kim}, \citenamefont {Ji}, \citenamefont {Daudlin}, \citenamefont {Okawachi}, \citenamefont {Cheng}, \citenamefont {Lipson}, \citenamefont {Gaeta} \emph {et~al.}}]{rizzo2023massively}%
  \BibitemOpen
  \bibfield  {author} {\bibinfo {author} {\bibfnamefont {A.}~\bibnamefont {Rizzo}}, \bibinfo {author} {\bibfnamefont {A.}~\bibnamefont {Novick}}, \bibinfo {author} {\bibfnamefont {V.}~\bibnamefont {Gopal}}, \bibinfo {author} {\bibfnamefont {B.~Y.}\ \bibnamefont {Kim}}, \bibinfo {author} {\bibfnamefont {X.}~\bibnamefont {Ji}}, \bibinfo {author} {\bibfnamefont {S.}~\bibnamefont {Daudlin}}, \bibinfo {author} {\bibfnamefont {Y.}~\bibnamefont {Okawachi}}, \bibinfo {author} {\bibfnamefont {Q.}~\bibnamefont {Cheng}}, \bibinfo {author} {\bibfnamefont {M.}~\bibnamefont {Lipson}}, \bibinfo {author} {\bibfnamefont {A.~L.}\ \bibnamefont {Gaeta}}, \emph {et~al.},\ }\bibfield  {title} {\bibinfo {title} {Massively scalable kerr comb-driven silicon photonic link},\ }\href@noop {} {\bibfield  {journal} {\bibinfo  {journal} {Nature Photonics}\ }\textbf {\bibinfo {volume} {17}},\ \bibinfo {pages} {781} (\bibinfo {year} {2023})}\BibitemShut {NoStop}%
\bibitem [{\citenamefont {Daudlin}\ \emph {et~al.}(2025)\citenamefont {Daudlin}, \citenamefont {Rizzo}, \citenamefont {Lee}, \citenamefont {Khilwani}, \citenamefont {Ou}, \citenamefont {Wang}, \citenamefont {Novick}, \citenamefont {Gopal}, \citenamefont {Cullen}, \citenamefont {Parsons} \emph {et~al.}}]{daudlin2025three}%
  \BibitemOpen
  \bibfield  {author} {\bibinfo {author} {\bibfnamefont {S.}~\bibnamefont {Daudlin}}, \bibinfo {author} {\bibfnamefont {A.}~\bibnamefont {Rizzo}}, \bibinfo {author} {\bibfnamefont {S.}~\bibnamefont {Lee}}, \bibinfo {author} {\bibfnamefont {D.}~\bibnamefont {Khilwani}}, \bibinfo {author} {\bibfnamefont {C.}~\bibnamefont {Ou}}, \bibinfo {author} {\bibfnamefont {S.}~\bibnamefont {Wang}}, \bibinfo {author} {\bibfnamefont {A.}~\bibnamefont {Novick}}, \bibinfo {author} {\bibfnamefont {V.}~\bibnamefont {Gopal}}, \bibinfo {author} {\bibfnamefont {M.}~\bibnamefont {Cullen}}, \bibinfo {author} {\bibfnamefont {R.}~\bibnamefont {Parsons}}, \emph {et~al.},\ }\bibfield  {title} {\bibinfo {title} {Three-dimensional photonic integration for ultra-low-energy, high-bandwidth interchip data links},\ }\href@noop {} {\bibfield  {journal} {\bibinfo  {journal} {Nature Photonics}\ ,\ \bibinfo {pages} {1}} (\bibinfo {year} {2025})}\BibitemShut {NoStop}%
\bibitem [{\citenamefont {Yuan}\ \emph {et~al.}(2024)\citenamefont {Yuan}, \citenamefont {Peng}, \citenamefont {Sorin}, \citenamefont {Cheung}, \citenamefont {Huang}, \citenamefont {Liang}, \citenamefont {Fiorentino},\ and\ \citenamefont {Beausoleil}}]{yuan20245}%
  \BibitemOpen
  \bibfield  {author} {\bibinfo {author} {\bibfnamefont {Y.}~\bibnamefont {Yuan}}, \bibinfo {author} {\bibfnamefont {Y.}~\bibnamefont {Peng}}, \bibinfo {author} {\bibfnamefont {W.~V.}\ \bibnamefont {Sorin}}, \bibinfo {author} {\bibfnamefont {S.}~\bibnamefont {Cheung}}, \bibinfo {author} {\bibfnamefont {Z.}~\bibnamefont {Huang}}, \bibinfo {author} {\bibfnamefont {D.}~\bibnamefont {Liang}}, \bibinfo {author} {\bibfnamefont {M.}~\bibnamefont {Fiorentino}},\ and\ \bibinfo {author} {\bibfnamefont {R.~G.}\ \bibnamefont {Beausoleil}},\ }\bibfield  {title} {\bibinfo {title} {A 5$\times$ 200 gbps microring modulator silicon chip empowered by two-segment z-shape junctions},\ }\href@noop {} {\bibfield  {journal} {\bibinfo  {journal} {Nature Communications}\ }\textbf {\bibinfo {volume} {15}},\ \bibinfo {pages} {918} (\bibinfo {year} {2024})}\BibitemShut {NoStop}%
\bibitem [{\citenamefont {Rizzo}\ \emph {et~al.}(2022)\citenamefont {Rizzo}, \citenamefont {Daudlin}, \citenamefont {Novick}, \citenamefont {James}, \citenamefont {Gopal}, \citenamefont {Murthy}, \citenamefont {Cheng}, \citenamefont {Kim}, \citenamefont {Ji}, \citenamefont {Okawachi} \emph {et~al.}}]{rizzo2022petabit}%
  \BibitemOpen
  \bibfield  {author} {\bibinfo {author} {\bibfnamefont {A.}~\bibnamefont {Rizzo}}, \bibinfo {author} {\bibfnamefont {S.}~\bibnamefont {Daudlin}}, \bibinfo {author} {\bibfnamefont {A.}~\bibnamefont {Novick}}, \bibinfo {author} {\bibfnamefont {A.}~\bibnamefont {James}}, \bibinfo {author} {\bibfnamefont {V.}~\bibnamefont {Gopal}}, \bibinfo {author} {\bibfnamefont {V.}~\bibnamefont {Murthy}}, \bibinfo {author} {\bibfnamefont {Q.}~\bibnamefont {Cheng}}, \bibinfo {author} {\bibfnamefont {B.~Y.}\ \bibnamefont {Kim}}, \bibinfo {author} {\bibfnamefont {X.}~\bibnamefont {Ji}}, \bibinfo {author} {\bibfnamefont {Y.}~\bibnamefont {Okawachi}}, \emph {et~al.},\ }\bibfield  {title} {\bibinfo {title} {Petabit-scale silicon photonic interconnects with integrated kerr frequency combs},\ }\href@noop {} {\bibfield  {journal} {\bibinfo  {journal} {IEEE Journal of Selected Topics in Quantum Electronics}\ }\textbf {\bibinfo {volume} {29}},\ \bibinfo {pages} {1} (\bibinfo {year} {2022})}\BibitemShut {NoStop}%
\bibitem [{\citenamefont {Wade}\ \emph {et~al.}(2018)\citenamefont {Wade}, \citenamefont {Davenport}, \citenamefont {Falco}, \citenamefont {Bhargava}, \citenamefont {Fini}, \citenamefont {Van~Orden}, \citenamefont {Meade}, \citenamefont {Yeung}, \citenamefont {Ram}, \citenamefont {Popovi{\'c}} \emph {et~al.}}]{wade2018bandwidth}%
  \BibitemOpen
  \bibfield  {author} {\bibinfo {author} {\bibfnamefont {M.}~\bibnamefont {Wade}}, \bibinfo {author} {\bibfnamefont {M.}~\bibnamefont {Davenport}}, \bibinfo {author} {\bibfnamefont {M.~D.~C.}\ \bibnamefont {Falco}}, \bibinfo {author} {\bibfnamefont {P.}~\bibnamefont {Bhargava}}, \bibinfo {author} {\bibfnamefont {J.}~\bibnamefont {Fini}}, \bibinfo {author} {\bibfnamefont {D.}~\bibnamefont {Van~Orden}}, \bibinfo {author} {\bibfnamefont {R.}~\bibnamefont {Meade}}, \bibinfo {author} {\bibfnamefont {E.}~\bibnamefont {Yeung}}, \bibinfo {author} {\bibfnamefont {R.}~\bibnamefont {Ram}}, \bibinfo {author} {\bibfnamefont {M.}~\bibnamefont {Popovi{\'c}}}, \emph {et~al.},\ }\bibfield  {title} {\bibinfo {title} {A bandwidth-dense, low power electronic-photonic platform and architecture for multi-tbps optical i/o},\ }in\ \href@noop {} {\emph {\bibinfo {booktitle} {2018 European Conference on Optical Communication (ECOC)}}}\ (\bibinfo {organization} {IEEE},\ \bibinfo {year} {2018})\ pp.\ \bibinfo {pages} {1--3}\BibitemShut
  {NoStop}%
\bibitem [{\citenamefont {Sun}\ \emph {et~al.}(2020)\citenamefont {Sun}, \citenamefont {Jeong}, \citenamefont {Zhang}, \citenamefont {Bae}, \citenamefont {Zhang}, \citenamefont {Bhargava}, \citenamefont {Van~Orden}, \citenamefont {Ardalan}, \citenamefont {Ramamurthy}, \citenamefont {Anderson} \emph {et~al.}}]{sun2020teraphy}%
  \BibitemOpen
  \bibfield  {author} {\bibinfo {author} {\bibfnamefont {C.}~\bibnamefont {Sun}}, \bibinfo {author} {\bibfnamefont {D.}~\bibnamefont {Jeong}}, \bibinfo {author} {\bibfnamefont {M.}~\bibnamefont {Zhang}}, \bibinfo {author} {\bibfnamefont {W.}~\bibnamefont {Bae}}, \bibinfo {author} {\bibfnamefont {C.}~\bibnamefont {Zhang}}, \bibinfo {author} {\bibfnamefont {P.}~\bibnamefont {Bhargava}}, \bibinfo {author} {\bibfnamefont {D.}~\bibnamefont {Van~Orden}}, \bibinfo {author} {\bibfnamefont {S.}~\bibnamefont {Ardalan}}, \bibinfo {author} {\bibfnamefont {C.}~\bibnamefont {Ramamurthy}}, \bibinfo {author} {\bibfnamefont {E.}~\bibnamefont {Anderson}}, \emph {et~al.},\ }\bibfield  {title} {\bibinfo {title} {Teraphy: An o-band wdm electro-optic platform for low power, terabit/s optical i/o},\ }in\ \href@noop {} {\emph {\bibinfo {booktitle} {2020 IEEE Symposium on VLSI Technology}}}\ (\bibinfo {organization} {IEEE},\ \bibinfo {year} {2020})\ pp.\ \bibinfo {pages} {1--2}\BibitemShut {NoStop}%
\bibitem [{\citenamefont {Sun}\ \emph {et~al.}(2015)\citenamefont {Sun}, \citenamefont {Wade}, \citenamefont {Lee}, \citenamefont {Orcutt}, \citenamefont {Alloatti}, \citenamefont {Georgas}, \citenamefont {Waterman}, \citenamefont {Shainline}, \citenamefont {Avizienis}, \citenamefont {Lin} \emph {et~al.}}]{sun2015single}%
  \BibitemOpen
  \bibfield  {author} {\bibinfo {author} {\bibfnamefont {C.}~\bibnamefont {Sun}}, \bibinfo {author} {\bibfnamefont {M.~T.}\ \bibnamefont {Wade}}, \bibinfo {author} {\bibfnamefont {Y.}~\bibnamefont {Lee}}, \bibinfo {author} {\bibfnamefont {J.~S.}\ \bibnamefont {Orcutt}}, \bibinfo {author} {\bibfnamefont {L.}~\bibnamefont {Alloatti}}, \bibinfo {author} {\bibfnamefont {M.~S.}\ \bibnamefont {Georgas}}, \bibinfo {author} {\bibfnamefont {A.~S.}\ \bibnamefont {Waterman}}, \bibinfo {author} {\bibfnamefont {J.~M.}\ \bibnamefont {Shainline}}, \bibinfo {author} {\bibfnamefont {R.~R.}\ \bibnamefont {Avizienis}}, \bibinfo {author} {\bibfnamefont {S.}~\bibnamefont {Lin}}, \emph {et~al.},\ }\bibfield  {title} {\bibinfo {title} {Single-chip microprocessor that communicates directly using light},\ }\href@noop {} {\bibfield  {journal} {\bibinfo  {journal} {Nature}\ }\textbf {\bibinfo {volume} {528}},\ \bibinfo {pages} {534} (\bibinfo {year} {2015})}\BibitemShut {NoStop}%
\bibitem [{\citenamefont {Atabaki}\ \emph {et~al.}(2018)\citenamefont {Atabaki}, \citenamefont {Moazeni}, \citenamefont {Pavanello}, \citenamefont {Gevorgyan}, \citenamefont {Notaros}, \citenamefont {Alloatti}, \citenamefont {Wade}, \citenamefont {Sun}, \citenamefont {Kruger}, \citenamefont {Meng} \emph {et~al.}}]{atabaki2018integrating}%
  \BibitemOpen
  \bibfield  {author} {\bibinfo {author} {\bibfnamefont {A.~H.}\ \bibnamefont {Atabaki}}, \bibinfo {author} {\bibfnamefont {S.}~\bibnamefont {Moazeni}}, \bibinfo {author} {\bibfnamefont {F.}~\bibnamefont {Pavanello}}, \bibinfo {author} {\bibfnamefont {H.}~\bibnamefont {Gevorgyan}}, \bibinfo {author} {\bibfnamefont {J.}~\bibnamefont {Notaros}}, \bibinfo {author} {\bibfnamefont {L.}~\bibnamefont {Alloatti}}, \bibinfo {author} {\bibfnamefont {M.~T.}\ \bibnamefont {Wade}}, \bibinfo {author} {\bibfnamefont {C.}~\bibnamefont {Sun}}, \bibinfo {author} {\bibfnamefont {S.~A.}\ \bibnamefont {Kruger}}, \bibinfo {author} {\bibfnamefont {H.}~\bibnamefont {Meng}}, \emph {et~al.},\ }\bibfield  {title} {\bibinfo {title} {Integrating photonics with silicon nanoelectronics for the next generation of systems on a chip},\ }\href@noop {} {\bibfield  {journal} {\bibinfo  {journal} {Nature}\ }\textbf {\bibinfo {volume} {556}},\ \bibinfo {pages} {349} (\bibinfo {year} {2018})}\BibitemShut {NoStop}%
\bibitem [{\citenamefont {Wade}\ \emph {et~al.}(2021)\citenamefont {Wade}, \citenamefont {Anderson}, \citenamefont {Ardalan}, \citenamefont {Bae}, \citenamefont {Beheshtian}, \citenamefont {Buchbinder}, \citenamefont {Chang}, \citenamefont {Chao}, \citenamefont {Eachempatti}, \citenamefont {Frey} \emph {et~al.}}]{wade2021error}%
  \BibitemOpen
  \bibfield  {author} {\bibinfo {author} {\bibfnamefont {M.}~\bibnamefont {Wade}}, \bibinfo {author} {\bibfnamefont {E.}~\bibnamefont {Anderson}}, \bibinfo {author} {\bibfnamefont {S.}~\bibnamefont {Ardalan}}, \bibinfo {author} {\bibfnamefont {W.}~\bibnamefont {Bae}}, \bibinfo {author} {\bibfnamefont {B.}~\bibnamefont {Beheshtian}}, \bibinfo {author} {\bibfnamefont {S.}~\bibnamefont {Buchbinder}}, \bibinfo {author} {\bibfnamefont {K.}~\bibnamefont {Chang}}, \bibinfo {author} {\bibfnamefont {P.}~\bibnamefont {Chao}}, \bibinfo {author} {\bibfnamefont {H.}~\bibnamefont {Eachempatti}}, \bibinfo {author} {\bibfnamefont {J.}~\bibnamefont {Frey}}, \emph {et~al.},\ }\bibfield  {title} {\bibinfo {title} {An error-free 1 tbps wdm optical i/o chiplet and multi-wavelength multi-port laser},\ }in\ \href@noop {} {\emph {\bibinfo {booktitle} {Optical Fiber Communication Conference}}}\ (\bibinfo {organization} {Optica Publishing Group},\ \bibinfo {year} {2021})\ pp.\ \bibinfo {pages} {F3C--6}\BibitemShut {NoStop}%
\bibitem [{\citenamefont {Beausoleil}\ \emph {et~al.}(2012)\citenamefont {Beausoleil}, \citenamefont {McLaren},\ and\ \citenamefont {Jouppi}}]{beausoleil2012photonic}%
  \BibitemOpen
  \bibfield  {author} {\bibinfo {author} {\bibfnamefont {R.~G.}\ \bibnamefont {Beausoleil}}, \bibinfo {author} {\bibfnamefont {M.}~\bibnamefont {McLaren}},\ and\ \bibinfo {author} {\bibfnamefont {N.~P.}\ \bibnamefont {Jouppi}},\ }\bibfield  {title} {\bibinfo {title} {Photonic architectures for high-performance data centers},\ }\href@noop {} {\bibfield  {journal} {\bibinfo  {journal} {IEEE Journal of Selected Topics in Quantum Electronics}\ }\textbf {\bibinfo {volume} {19}},\ \bibinfo {pages} {3700109} (\bibinfo {year} {2012})}\BibitemShut {NoStop}%
\bibitem [{\citenamefont {Miller}(1997)}]{miller1997physical}%
  \BibitemOpen
  \bibfield  {author} {\bibinfo {author} {\bibfnamefont {D.~A.}\ \bibnamefont {Miller}},\ }\bibfield  {title} {\bibinfo {title} {Physical reasons for optical interconnection},\ }\href@noop {} {\bibfield  {journal} {\bibinfo  {journal} {Intel J. Optoelectronics}\ }\textbf {\bibinfo {volume} {11}},\ \bibinfo {pages} {155} (\bibinfo {year} {1997})}\BibitemShut {NoStop}%
\bibitem [{\citenamefont {Timurdogan}\ \emph {et~al.}(2014)\citenamefont {Timurdogan}, \citenamefont {Sorace-Agaskar}, \citenamefont {Sun}, \citenamefont {Shah~Hosseini}, \citenamefont {Biberman},\ and\ \citenamefont {Watts}}]{timurdogan2014ultralow}%
  \BibitemOpen
  \bibfield  {author} {\bibinfo {author} {\bibfnamefont {E.}~\bibnamefont {Timurdogan}}, \bibinfo {author} {\bibfnamefont {C.~M.}\ \bibnamefont {Sorace-Agaskar}}, \bibinfo {author} {\bibfnamefont {J.}~\bibnamefont {Sun}}, \bibinfo {author} {\bibfnamefont {E.}~\bibnamefont {Shah~Hosseini}}, \bibinfo {author} {\bibfnamefont {A.}~\bibnamefont {Biberman}},\ and\ \bibinfo {author} {\bibfnamefont {M.~R.}\ \bibnamefont {Watts}},\ }\bibfield  {title} {\bibinfo {title} {An ultralow power athermal silicon modulator},\ }\href@noop {} {\bibfield  {journal} {\bibinfo  {journal} {Nature communications}\ }\textbf {\bibinfo {volume} {5}},\ \bibinfo {pages} {1} (\bibinfo {year} {2014})}\BibitemShut {NoStop}%
\bibitem [{\citenamefont {Shekhar}\ \emph {et~al.}(2024)\citenamefont {Shekhar}, \citenamefont {Bogaerts}, \citenamefont {Chrostowski}, \citenamefont {Bowers}, \citenamefont {Hochberg}, \citenamefont {Soref},\ and\ \citenamefont {Shastri}}]{shekhar2024roadmapping}%
  \BibitemOpen
  \bibfield  {author} {\bibinfo {author} {\bibfnamefont {S.}~\bibnamefont {Shekhar}}, \bibinfo {author} {\bibfnamefont {W.}~\bibnamefont {Bogaerts}}, \bibinfo {author} {\bibfnamefont {L.}~\bibnamefont {Chrostowski}}, \bibinfo {author} {\bibfnamefont {J.~E.}\ \bibnamefont {Bowers}}, \bibinfo {author} {\bibfnamefont {M.}~\bibnamefont {Hochberg}}, \bibinfo {author} {\bibfnamefont {R.}~\bibnamefont {Soref}},\ and\ \bibinfo {author} {\bibfnamefont {B.~J.}\ \bibnamefont {Shastri}},\ }\bibfield  {title} {\bibinfo {title} {Roadmapping the next generation of silicon photonics},\ }\href@noop {} {\bibfield  {journal} {\bibinfo  {journal} {Nature Communications}\ }\textbf {\bibinfo {volume} {15}},\ \bibinfo {pages} {751} (\bibinfo {year} {2024})}\BibitemShut {NoStop}%
\bibitem [{\citenamefont {Rahim}\ \emph {et~al.}(2021)\citenamefont {Rahim}, \citenamefont {Hermans}, \citenamefont {Wohlfeil}, \citenamefont {Petousi}, \citenamefont {Kuyken}, \citenamefont {Van~Thourhout},\ and\ \citenamefont {Baets}}]{rahim2021taking}%
  \BibitemOpen
  \bibfield  {author} {\bibinfo {author} {\bibfnamefont {A.}~\bibnamefont {Rahim}}, \bibinfo {author} {\bibfnamefont {A.}~\bibnamefont {Hermans}}, \bibinfo {author} {\bibfnamefont {B.}~\bibnamefont {Wohlfeil}}, \bibinfo {author} {\bibfnamefont {D.}~\bibnamefont {Petousi}}, \bibinfo {author} {\bibfnamefont {B.}~\bibnamefont {Kuyken}}, \bibinfo {author} {\bibfnamefont {D.}~\bibnamefont {Van~Thourhout}},\ and\ \bibinfo {author} {\bibfnamefont {R.}~\bibnamefont {Baets}},\ }\bibfield  {title} {\bibinfo {title} {Taking silicon photonics modulators to a higher performance level: state-of-the-art and a review of new technologies},\ }\href@noop {} {\bibfield  {journal} {\bibinfo  {journal} {Advanced Photonics}\ }\textbf {\bibinfo {volume} {3}},\ \bibinfo {pages} {024003} (\bibinfo {year} {2021})}\BibitemShut {NoStop}%
\bibitem [{\citenamefont {Li}\ \emph {et~al.}(2016)\citenamefont {Li}, \citenamefont {Zheng}, \citenamefont {Krishnamoorthy},\ and\ \citenamefont {Buckwalter}}]{li2016scaling}%
  \BibitemOpen
  \bibfield  {author} {\bibinfo {author} {\bibfnamefont {J.}~\bibnamefont {Li}}, \bibinfo {author} {\bibfnamefont {X.}~\bibnamefont {Zheng}}, \bibinfo {author} {\bibfnamefont {A.~V.}\ \bibnamefont {Krishnamoorthy}},\ and\ \bibinfo {author} {\bibfnamefont {J.~F.}\ \bibnamefont {Buckwalter}},\ }\bibfield  {title} {\bibinfo {title} {Scaling trends for picojoule-per-bit wdm photonic interconnects in cmos soi and finfet processes},\ }\href@noop {} {\bibfield  {journal} {\bibinfo  {journal} {Journal of Lightwave Technology}\ }\textbf {\bibinfo {volume} {34}},\ \bibinfo {pages} {2730} (\bibinfo {year} {2016})}\BibitemShut {NoStop}%
\bibitem [{\citenamefont {Jalali}\ and\ \citenamefont {Fathpour}(2006)}]{jalali2006silicon}%
  \BibitemOpen
  \bibfield  {author} {\bibinfo {author} {\bibfnamefont {B.}~\bibnamefont {Jalali}}\ and\ \bibinfo {author} {\bibfnamefont {S.}~\bibnamefont {Fathpour}},\ }\bibfield  {title} {\bibinfo {title} {Silicon photonics},\ }\href@noop {} {\bibfield  {journal} {\bibinfo  {journal} {Journal of lightwave technology}\ }\textbf {\bibinfo {volume} {24}},\ \bibinfo {pages} {4600} (\bibinfo {year} {2006})}\BibitemShut {NoStop}%
\bibitem [{\citenamefont {Soref}(2007)}]{soref2007past}%
  \BibitemOpen
  \bibfield  {author} {\bibinfo {author} {\bibfnamefont {R.}~\bibnamefont {Soref}},\ }\bibfield  {title} {\bibinfo {title} {The past, present, and future of silicon photonics},\ }\href@noop {} {\bibfield  {journal} {\bibinfo  {journal} {IEEE Journal of selected topics in quantum electronics}\ }\textbf {\bibinfo {volume} {12}},\ \bibinfo {pages} {1678} (\bibinfo {year} {2007})}\BibitemShut {NoStop}%
\bibitem [{\citenamefont {Siew}\ \emph {et~al.}(2021)\citenamefont {Siew}, \citenamefont {Li}, \citenamefont {Gao}, \citenamefont {Zheng}, \citenamefont {Zhang}, \citenamefont {Guo}, \citenamefont {Xie}, \citenamefont {Song}, \citenamefont {Dong}, \citenamefont {Luo} \emph {et~al.}}]{siew2021review}%
  \BibitemOpen
  \bibfield  {author} {\bibinfo {author} {\bibfnamefont {S.~Y.}\ \bibnamefont {Siew}}, \bibinfo {author} {\bibfnamefont {B.}~\bibnamefont {Li}}, \bibinfo {author} {\bibfnamefont {F.}~\bibnamefont {Gao}}, \bibinfo {author} {\bibfnamefont {H.~Y.}\ \bibnamefont {Zheng}}, \bibinfo {author} {\bibfnamefont {W.}~\bibnamefont {Zhang}}, \bibinfo {author} {\bibfnamefont {P.}~\bibnamefont {Guo}}, \bibinfo {author} {\bibfnamefont {S.~W.}\ \bibnamefont {Xie}}, \bibinfo {author} {\bibfnamefont {A.}~\bibnamefont {Song}}, \bibinfo {author} {\bibfnamefont {B.}~\bibnamefont {Dong}}, \bibinfo {author} {\bibfnamefont {L.~W.}\ \bibnamefont {Luo}}, \emph {et~al.},\ }\bibfield  {title} {\bibinfo {title} {Review of silicon photonics technology and platform development},\ }\href@noop {} {\bibfield  {journal} {\bibinfo  {journal} {Journal of Lightwave Technology}\ }\textbf {\bibinfo {volume} {39}},\ \bibinfo {pages} {4374} (\bibinfo {year} {2021})}\BibitemShut {NoStop}%
\bibitem [{\citenamefont {Geravand}\ \emph {et~al.}(2025)\citenamefont {Geravand}, \citenamefont {Zheng}, \citenamefont {Shateri}, \citenamefont {Levasseur}, \citenamefont {Rusch},\ and\ \citenamefont {Shi}}]{geravand2025ultrafast}%
  \BibitemOpen
  \bibfield  {author} {\bibinfo {author} {\bibfnamefont {A.}~\bibnamefont {Geravand}}, \bibinfo {author} {\bibfnamefont {Z.}~\bibnamefont {Zheng}}, \bibinfo {author} {\bibfnamefont {F.}~\bibnamefont {Shateri}}, \bibinfo {author} {\bibfnamefont {S.}~\bibnamefont {Levasseur}}, \bibinfo {author} {\bibfnamefont {L.~A.}\ \bibnamefont {Rusch}},\ and\ \bibinfo {author} {\bibfnamefont {W.}~\bibnamefont {Shi}},\ }\bibfield  {title} {\bibinfo {title} {Ultrafast coherent dynamics of microring modulators},\ }\href@noop {} {\bibfield  {journal} {\bibinfo  {journal} {Nature Photonics}\ ,\ \bibinfo {pages} {1}} (\bibinfo {year} {2025})}\BibitemShut {NoStop}%
\bibitem [{\citenamefont {Bogaerts}\ \emph {et~al.}(2012)\citenamefont {Bogaerts}, \citenamefont {De~Heyn}, \citenamefont {Van~Vaerenbergh}, \citenamefont {De~Vos}, \citenamefont {Kumar~Selvaraja}, \citenamefont {Claes}, \citenamefont {Dumon}, \citenamefont {Bienstman}, \citenamefont {Van~Thourhout},\ and\ \citenamefont {Baets}}]{bogaerts2012silicon}%
  \BibitemOpen
  \bibfield  {author} {\bibinfo {author} {\bibfnamefont {W.}~\bibnamefont {Bogaerts}}, \bibinfo {author} {\bibfnamefont {P.}~\bibnamefont {De~Heyn}}, \bibinfo {author} {\bibfnamefont {T.}~\bibnamefont {Van~Vaerenbergh}}, \bibinfo {author} {\bibfnamefont {K.}~\bibnamefont {De~Vos}}, \bibinfo {author} {\bibfnamefont {S.}~\bibnamefont {Kumar~Selvaraja}}, \bibinfo {author} {\bibfnamefont {T.}~\bibnamefont {Claes}}, \bibinfo {author} {\bibfnamefont {P.}~\bibnamefont {Dumon}}, \bibinfo {author} {\bibfnamefont {P.}~\bibnamefont {Bienstman}}, \bibinfo {author} {\bibfnamefont {D.}~\bibnamefont {Van~Thourhout}},\ and\ \bibinfo {author} {\bibfnamefont {R.}~\bibnamefont {Baets}},\ }\bibfield  {title} {\bibinfo {title} {Silicon microring resonators},\ }\href@noop {} {\bibfield  {journal} {\bibinfo  {journal} {Laser \& Photonics Reviews}\ }\textbf {\bibinfo {volume} {6}},\ \bibinfo {pages} {47} (\bibinfo {year} {2012})}\BibitemShut {NoStop}%
\bibitem [{\citenamefont {Nikdast}\ \emph {et~al.}(2016)\citenamefont {Nikdast}, \citenamefont {Nicolescu}, \citenamefont {Trajkovic},\ and\ \citenamefont {Liboiron-Ladouceur}}]{nikdast2016chip}%
  \BibitemOpen
  \bibfield  {author} {\bibinfo {author} {\bibfnamefont {M.}~\bibnamefont {Nikdast}}, \bibinfo {author} {\bibfnamefont {G.}~\bibnamefont {Nicolescu}}, \bibinfo {author} {\bibfnamefont {J.}~\bibnamefont {Trajkovic}},\ and\ \bibinfo {author} {\bibfnamefont {O.}~\bibnamefont {Liboiron-Ladouceur}},\ }\bibfield  {title} {\bibinfo {title} {Chip-scale silicon photonic interconnects: A formal study on fabrication non-uniformity},\ }\href@noop {} {\bibfield  {journal} {\bibinfo  {journal} {Journal of Lightwave Technology}\ }\textbf {\bibinfo {volume} {34}},\ \bibinfo {pages} {3682} (\bibinfo {year} {2016})}\BibitemShut {NoStop}%
\bibitem [{\citenamefont {Pintus}\ \emph {et~al.}(2019)\citenamefont {Pintus}, \citenamefont {Hofbauer}, \citenamefont {Manganelli}, \citenamefont {Fournier}, \citenamefont {Gundavarapu}, \citenamefont {Lemonnier}, \citenamefont {Gambini}, \citenamefont {Adelmini}, \citenamefont {Meinhart}, \citenamefont {Kopp} \emph {et~al.}}]{pintus2019pwm}%
  \BibitemOpen
  \bibfield  {author} {\bibinfo {author} {\bibfnamefont {P.}~\bibnamefont {Pintus}}, \bibinfo {author} {\bibfnamefont {M.}~\bibnamefont {Hofbauer}}, \bibinfo {author} {\bibfnamefont {C.~L.}\ \bibnamefont {Manganelli}}, \bibinfo {author} {\bibfnamefont {M.}~\bibnamefont {Fournier}}, \bibinfo {author} {\bibfnamefont {S.}~\bibnamefont {Gundavarapu}}, \bibinfo {author} {\bibfnamefont {O.}~\bibnamefont {Lemonnier}}, \bibinfo {author} {\bibfnamefont {F.}~\bibnamefont {Gambini}}, \bibinfo {author} {\bibfnamefont {L.}~\bibnamefont {Adelmini}}, \bibinfo {author} {\bibfnamefont {C.}~\bibnamefont {Meinhart}}, \bibinfo {author} {\bibfnamefont {C.}~\bibnamefont {Kopp}}, \emph {et~al.},\ }\bibfield  {title} {\bibinfo {title} {Pwm-driven thermally tunable silicon microring resonators: design, fabrication, and characterization},\ }\href@noop {} {\bibfield  {journal} {\bibinfo  {journal} {Laser \& Photonics Reviews}\ }\textbf {\bibinfo {volume} {13}},\ \bibinfo {pages} {1800275} (\bibinfo {year} {2019})}\BibitemShut {NoStop}%
\bibitem [{\citenamefont {Jayatilleka}\ \emph {et~al.}(2021)\citenamefont {Jayatilleka}, \citenamefont {Frish}, \citenamefont {Kumar}, \citenamefont {Heck}, \citenamefont {Ma}, \citenamefont {Sakib}, \citenamefont {Huang},\ and\ \citenamefont {Rong}}]{jayatilleka2021post}%
  \BibitemOpen
  \bibfield  {author} {\bibinfo {author} {\bibfnamefont {H.}~\bibnamefont {Jayatilleka}}, \bibinfo {author} {\bibfnamefont {H.}~\bibnamefont {Frish}}, \bibinfo {author} {\bibfnamefont {R.}~\bibnamefont {Kumar}}, \bibinfo {author} {\bibfnamefont {J.}~\bibnamefont {Heck}}, \bibinfo {author} {\bibfnamefont {C.}~\bibnamefont {Ma}}, \bibinfo {author} {\bibfnamefont {M.~N.}\ \bibnamefont {Sakib}}, \bibinfo {author} {\bibfnamefont {D.}~\bibnamefont {Huang}},\ and\ \bibinfo {author} {\bibfnamefont {H.}~\bibnamefont {Rong}},\ }\bibfield  {title} {\bibinfo {title} {Post-fabrication trimming of silicon photonic ring resonators at wafer-scale},\ }\href@noop {} {\bibfield  {journal} {\bibinfo  {journal} {Journal of Lightwave Technology}\ }\textbf {\bibinfo {volume} {39}},\ \bibinfo {pages} {5083} (\bibinfo {year} {2021})}\BibitemShut {NoStop}%
\bibitem [{\citenamefont {Zhou}\ \emph {et~al.}(2009)\citenamefont {Zhou}, \citenamefont {Okamoto},\ and\ \citenamefont {Yoo}}]{zhou2009athermalizing}%
  \BibitemOpen
  \bibfield  {author} {\bibinfo {author} {\bibfnamefont {L.}~\bibnamefont {Zhou}}, \bibinfo {author} {\bibfnamefont {K.}~\bibnamefont {Okamoto}},\ and\ \bibinfo {author} {\bibfnamefont {S.~B.}\ \bibnamefont {Yoo}},\ }\bibfield  {title} {\bibinfo {title} {Athermalizing and trimming of slotted silicon microring resonators with uv-sensitive pmma upper-cladding},\ }\href@noop {} {\bibfield  {journal} {\bibinfo  {journal} {IEEE Photonics Technology Letters}\ }\textbf {\bibinfo {volume} {21}},\ \bibinfo {pages} {1175} (\bibinfo {year} {2009})}\BibitemShut {NoStop}%
\bibitem [{\citenamefont {Schrauwen}\ \emph {et~al.}(2008)\citenamefont {Schrauwen}, \citenamefont {Van~Thourhout},\ and\ \citenamefont {Baets}}]{schrauwen2008trimming}%
  \BibitemOpen
  \bibfield  {author} {\bibinfo {author} {\bibfnamefont {J.}~\bibnamefont {Schrauwen}}, \bibinfo {author} {\bibfnamefont {D.}~\bibnamefont {Van~Thourhout}},\ and\ \bibinfo {author} {\bibfnamefont {R.}~\bibnamefont {Baets}},\ }\bibfield  {title} {\bibinfo {title} {Trimming of silicon ring resonator by electron beam induced compaction and strain},\ }\href@noop {} {\bibfield  {journal} {\bibinfo  {journal} {Optics express}\ }\textbf {\bibinfo {volume} {16}},\ \bibinfo {pages} {3738} (\bibinfo {year} {2008})}\BibitemShut {NoStop}%
\bibitem [{\citenamefont {Chu}\ \emph {et~al.}(1999)\citenamefont {Chu}, \citenamefont {Pan}, \citenamefont {Sato}, \citenamefont {Kaneko}, \citenamefont {Little},\ and\ \citenamefont {Kokubun}}]{chu1999wavelength}%
  \BibitemOpen
  \bibfield  {author} {\bibinfo {author} {\bibfnamefont {S.~T.}\ \bibnamefont {Chu}}, \bibinfo {author} {\bibfnamefont {W.}~\bibnamefont {Pan}}, \bibinfo {author} {\bibfnamefont {S.}~\bibnamefont {Sato}}, \bibinfo {author} {\bibfnamefont {T.}~\bibnamefont {Kaneko}}, \bibinfo {author} {\bibfnamefont {B.~E.}\ \bibnamefont {Little}},\ and\ \bibinfo {author} {\bibfnamefont {Y.}~\bibnamefont {Kokubun}},\ }\bibfield  {title} {\bibinfo {title} {Wavelength trimming of a microring resonator filter by means of a uv sensitive polymer overlay},\ }\href@noop {} {\bibfield  {journal} {\bibinfo  {journal} {IEEE Photonics Technology Letters}\ }\textbf {\bibinfo {volume} {11}},\ \bibinfo {pages} {688} (\bibinfo {year} {1999})}\BibitemShut {NoStop}%
\bibitem [{\citenamefont {Wu}\ \emph {et~al.}(2025)\citenamefont {Wu}, \citenamefont {Sun}, \citenamefont {Xiong}, \citenamefont {Yv}, \citenamefont {Zhang}, \citenamefont {Zheng}, \citenamefont {Ma},\ and\ \citenamefont {Chu}}]{wu2025lossless}%
  \BibitemOpen
  \bibfield  {author} {\bibinfo {author} {\bibfnamefont {Y.}~\bibnamefont {Wu}}, \bibinfo {author} {\bibfnamefont {H.}~\bibnamefont {Sun}}, \bibinfo {author} {\bibfnamefont {B.}~\bibnamefont {Xiong}}, \bibinfo {author} {\bibfnamefont {Y.}~\bibnamefont {Yv}}, \bibinfo {author} {\bibfnamefont {J.}~\bibnamefont {Zhang}}, \bibinfo {author} {\bibfnamefont {Z.}~\bibnamefont {Zheng}}, \bibinfo {author} {\bibfnamefont {W.}~\bibnamefont {Ma}},\ and\ \bibinfo {author} {\bibfnamefont {T.}~\bibnamefont {Chu}},\ }\bibfield  {title} {\bibinfo {title} {Lossless, non-volatile post-fabrication trimming of pics via on-chip high-temperature annealing of undercut waveguides},\ }\href@noop {} {\bibfield  {journal} {\bibinfo  {journal} {arXiv preprint arXiv:2506.18633}\ } (\bibinfo {year} {2025})}\BibitemShut {NoStop}%
\bibitem [{\citenamefont {Chen}\ \emph {et~al.}(2025)\citenamefont {Chen}, \citenamefont {Xue}, \citenamefont {Yong}, \citenamefont {Luo}, \citenamefont {Chua}, \citenamefont {Stalmashonak}, \citenamefont {Lo}, \citenamefont {Poon},\ and\ \citenamefont {Sacher}}]{chen2025thermally}%
  \BibitemOpen
  \bibfield  {author} {\bibinfo {author} {\bibfnamefont {H.}~\bibnamefont {Chen}}, \bibinfo {author} {\bibfnamefont {T.}~\bibnamefont {Xue}}, \bibinfo {author} {\bibfnamefont {Z.}~\bibnamefont {Yong}}, \bibinfo {author} {\bibfnamefont {X.}~\bibnamefont {Luo}}, \bibinfo {author} {\bibfnamefont {H.}~\bibnamefont {Chua}}, \bibinfo {author} {\bibfnamefont {A.}~\bibnamefont {Stalmashonak}}, \bibinfo {author} {\bibfnamefont {G.-Q.}\ \bibnamefont {Lo}}, \bibinfo {author} {\bibfnamefont {J.~K.}\ \bibnamefont {Poon}},\ and\ \bibinfo {author} {\bibfnamefont {W.~D.}\ \bibnamefont {Sacher}},\ }\bibfield  {title} {\bibinfo {title} {Thermally induced refractive index trimming of visible-light silicon nitride waveguides using suspended heaters},\ }\href@noop {} {\bibfield  {journal} {\bibinfo  {journal} {arXiv preprint arXiv:2504.21262}\ } (\bibinfo {year} {2025})}\BibitemShut {NoStop}%
\bibitem [{\citenamefont {Farmakidis}\ \emph {et~al.}(2023)\citenamefont {Farmakidis}, \citenamefont {Yu}, \citenamefont {Lee}, \citenamefont {Feldmann}, \citenamefont {Wang}, \citenamefont {He}, \citenamefont {Aggarwal}, \citenamefont {Dong}, \citenamefont {Pernice},\ and\ \citenamefont {Bhaskaran}}]{farmakidis2023scalable}%
  \BibitemOpen
  \bibfield  {author} {\bibinfo {author} {\bibfnamefont {N.}~\bibnamefont {Farmakidis}}, \bibinfo {author} {\bibfnamefont {H.}~\bibnamefont {Yu}}, \bibinfo {author} {\bibfnamefont {J.~S.}\ \bibnamefont {Lee}}, \bibinfo {author} {\bibfnamefont {J.}~\bibnamefont {Feldmann}}, \bibinfo {author} {\bibfnamefont {M.}~\bibnamefont {Wang}}, \bibinfo {author} {\bibfnamefont {Y.}~\bibnamefont {He}}, \bibinfo {author} {\bibfnamefont {S.}~\bibnamefont {Aggarwal}}, \bibinfo {author} {\bibfnamefont {B.}~\bibnamefont {Dong}}, \bibinfo {author} {\bibfnamefont {W.~H.}\ \bibnamefont {Pernice}},\ and\ \bibinfo {author} {\bibfnamefont {H.}~\bibnamefont {Bhaskaran}},\ }\bibfield  {title} {\bibinfo {title} {Scalable high-precision trimming of photonic resonances by polymer exposure to energetic beams},\ }\href@noop {} {\bibfield  {journal} {\bibinfo  {journal} {Nano Letters}\ }\textbf {\bibinfo {volume} {23}},\ \bibinfo {pages} {4800} (\bibinfo {year} {2023})}\BibitemShut {NoStop}%
\bibitem [{\citenamefont {Menssen}\ \emph {et~al.}(2023)\citenamefont {Menssen}, \citenamefont {Hermans}, \citenamefont {Christen}, \citenamefont {Propson}, \citenamefont {Li}, \citenamefont {Leenheer}, \citenamefont {Zimmermann}, \citenamefont {Dong}, \citenamefont {Larocque}, \citenamefont {Raniwala}, \citenamefont {Gilbert}, \citenamefont {Eichenfield},\ and\ \citenamefont {Englund}}]{Menssen:23}%
  \BibitemOpen
  \bibfield  {author} {\bibinfo {author} {\bibfnamefont {A.~J.}\ \bibnamefont {Menssen}}, \bibinfo {author} {\bibfnamefont {A.}~\bibnamefont {Hermans}}, \bibinfo {author} {\bibfnamefont {I.}~\bibnamefont {Christen}}, \bibinfo {author} {\bibfnamefont {T.}~\bibnamefont {Propson}}, \bibinfo {author} {\bibfnamefont {C.}~\bibnamefont {Li}}, \bibinfo {author} {\bibfnamefont {A.~J.}\ \bibnamefont {Leenheer}}, \bibinfo {author} {\bibfnamefont {M.}~\bibnamefont {Zimmermann}}, \bibinfo {author} {\bibfnamefont {M.}~\bibnamefont {Dong}}, \bibinfo {author} {\bibfnamefont {H.}~\bibnamefont {Larocque}}, \bibinfo {author} {\bibfnamefont {H.}~\bibnamefont {Raniwala}}, \bibinfo {author} {\bibfnamefont {G.}~\bibnamefont {Gilbert}}, \bibinfo {author} {\bibfnamefont {M.}~\bibnamefont {Eichenfield}},\ and\ \bibinfo {author} {\bibfnamefont {D.~R.}\ \bibnamefont {Englund}},\ }\bibfield  {title} {\bibinfo {title} {Scalable photonic integrated circuits for high-fidelity light control},\ }\href {https://doi.org/10.1364/OPTICA.489504}
  {\bibfield  {journal} {\bibinfo  {journal} {Optica}\ }\textbf {\bibinfo {volume} {10}},\ \bibinfo {pages} {1366} (\bibinfo {year} {2023})}\BibitemShut {NoStop}%
\bibitem [{\citenamefont {Butt}\ \emph {et~al.}(2025)\citenamefont {Butt}, \citenamefont {Imran~Akca},\ and\ \citenamefont {Mateos}}]{butt2025integrated}%
  \BibitemOpen
  \bibfield  {author} {\bibinfo {author} {\bibfnamefont {M.~A.}\ \bibnamefont {Butt}}, \bibinfo {author} {\bibfnamefont {B.}~\bibnamefont {Imran~Akca}},\ and\ \bibinfo {author} {\bibfnamefont {X.}~\bibnamefont {Mateos}},\ }\bibfield  {title} {\bibinfo {title} {Integrated photonic biosensors: Enabling next-generation lab-on-a-chip platforms},\ }\href@noop {} {\bibfield  {journal} {\bibinfo  {journal} {Nanomaterials}\ }\textbf {\bibinfo {volume} {15}},\ \bibinfo {pages} {731} (\bibinfo {year} {2025})}\BibitemShut {NoStop}%
\bibitem [{\citenamefont {Kazanskiy}\ \emph {et~al.}(2023)\citenamefont {Kazanskiy}, \citenamefont {Khonina},\ and\ \citenamefont {Butt}}]{kazanskiy2023review}%
  \BibitemOpen
  \bibfield  {author} {\bibinfo {author} {\bibfnamefont {N.~L.}\ \bibnamefont {Kazanskiy}}, \bibinfo {author} {\bibfnamefont {S.~N.}\ \bibnamefont {Khonina}},\ and\ \bibinfo {author} {\bibfnamefont {M.~A.}\ \bibnamefont {Butt}},\ }\bibfield  {title} {\bibinfo {title} {A review of photonic sensors based on ring resonator structures: three widely used platforms and implications of sensing applications},\ }\href@noop {} {\bibfield  {journal} {\bibinfo  {journal} {Micromachines}\ }\textbf {\bibinfo {volume} {14}},\ \bibinfo {pages} {1080} (\bibinfo {year} {2023})}\BibitemShut {NoStop}%
\bibitem [{\citenamefont {Chakraborty}\ \emph {et~al.}(2020)\citenamefont {Chakraborty}, \citenamefont {Carolan}, \citenamefont {Clark}, \citenamefont {Bunandar}, \citenamefont {Gilbert}, \citenamefont {Notaros}, \citenamefont {Watts},\ and\ \citenamefont {Englund}}]{chakraborty2020cryogenic}%
  \BibitemOpen
  \bibfield  {author} {\bibinfo {author} {\bibfnamefont {U.}~\bibnamefont {Chakraborty}}, \bibinfo {author} {\bibfnamefont {J.}~\bibnamefont {Carolan}}, \bibinfo {author} {\bibfnamefont {G.}~\bibnamefont {Clark}}, \bibinfo {author} {\bibfnamefont {D.}~\bibnamefont {Bunandar}}, \bibinfo {author} {\bibfnamefont {G.}~\bibnamefont {Gilbert}}, \bibinfo {author} {\bibfnamefont {J.}~\bibnamefont {Notaros}}, \bibinfo {author} {\bibfnamefont {M.~R.}\ \bibnamefont {Watts}},\ and\ \bibinfo {author} {\bibfnamefont {D.~R.}\ \bibnamefont {Englund}},\ }\bibfield  {title} {\bibinfo {title} {Cryogenic operation of silicon photonic modulators based on the dc kerr effect},\ }\href@noop {} {\bibfield  {journal} {\bibinfo  {journal} {Optica}\ }\textbf {\bibinfo {volume} {7}},\ \bibinfo {pages} {1385} (\bibinfo {year} {2020})}\BibitemShut {NoStop}%
\bibitem [{\citenamefont {Reed}\ \emph {et~al.}(2010)\citenamefont {Reed}, \citenamefont {Mashanovich}, \citenamefont {Gardes},\ and\ \citenamefont {Thomson}}]{reed2010silicon}%
  \BibitemOpen
  \bibfield  {author} {\bibinfo {author} {\bibfnamefont {G.~T.}\ \bibnamefont {Reed}}, \bibinfo {author} {\bibfnamefont {G.}~\bibnamefont {Mashanovich}}, \bibinfo {author} {\bibfnamefont {F.~Y.}\ \bibnamefont {Gardes}},\ and\ \bibinfo {author} {\bibfnamefont {D.}~\bibnamefont {Thomson}},\ }\bibfield  {title} {\bibinfo {title} {Silicon optical modulators},\ }\href@noop {} {\bibfield  {journal} {\bibinfo  {journal} {Nature photonics}\ }\textbf {\bibinfo {volume} {4}},\ \bibinfo {pages} {518} (\bibinfo {year} {2010})}\BibitemShut {NoStop}%
\bibitem [{\citenamefont {Xu}\ \emph {et~al.}(2005)\citenamefont {Xu}, \citenamefont {Schmidt}, \citenamefont {Pradhan},\ and\ \citenamefont {Lipson}}]{xu2005micrometre}%
  \BibitemOpen
  \bibfield  {author} {\bibinfo {author} {\bibfnamefont {Q.}~\bibnamefont {Xu}}, \bibinfo {author} {\bibfnamefont {B.}~\bibnamefont {Schmidt}}, \bibinfo {author} {\bibfnamefont {S.}~\bibnamefont {Pradhan}},\ and\ \bibinfo {author} {\bibfnamefont {M.}~\bibnamefont {Lipson}},\ }\bibfield  {title} {\bibinfo {title} {Micrometre-scale silicon electro-optic modulator},\ }\href@noop {} {\bibfield  {journal} {\bibinfo  {journal} {nature}\ }\textbf {\bibinfo {volume} {435}},\ \bibinfo {pages} {325} (\bibinfo {year} {2005})}\BibitemShut {NoStop}%
\bibitem [{\citenamefont {Witzens}(2018)}]{witzens2018high}%
  \BibitemOpen
  \bibfield  {author} {\bibinfo {author} {\bibfnamefont {J.}~\bibnamefont {Witzens}},\ }\bibfield  {title} {\bibinfo {title} {High-speed silicon photonics modulators},\ }\href@noop {} {\bibfield  {journal} {\bibinfo  {journal} {Proceedings of the IEEE}\ }\textbf {\bibinfo {volume} {106}},\ \bibinfo {pages} {2158} (\bibinfo {year} {2018})}\BibitemShut {NoStop}%
\bibitem [{\citenamefont {Liu}\ \emph {et~al.}(2004)\citenamefont {Liu}, \citenamefont {Jones}, \citenamefont {Liao}, \citenamefont {Samara-Rubio}, \citenamefont {Rubin}, \citenamefont {Cohen}, \citenamefont {Nicolaescu},\ and\ \citenamefont {Paniccia}}]{liu2004high}%
  \BibitemOpen
  \bibfield  {author} {\bibinfo {author} {\bibfnamefont {A.}~\bibnamefont {Liu}}, \bibinfo {author} {\bibfnamefont {R.}~\bibnamefont {Jones}}, \bibinfo {author} {\bibfnamefont {L.}~\bibnamefont {Liao}}, \bibinfo {author} {\bibfnamefont {D.}~\bibnamefont {Samara-Rubio}}, \bibinfo {author} {\bibfnamefont {D.}~\bibnamefont {Rubin}}, \bibinfo {author} {\bibfnamefont {O.}~\bibnamefont {Cohen}}, \bibinfo {author} {\bibfnamefont {R.}~\bibnamefont {Nicolaescu}},\ and\ \bibinfo {author} {\bibfnamefont {M.}~\bibnamefont {Paniccia}},\ }\bibfield  {title} {\bibinfo {title} {A high-speed silicon optical modulator based on a metal--oxide--semiconductor capacitor},\ }\href@noop {} {\bibfield  {journal} {\bibinfo  {journal} {Nature}\ }\textbf {\bibinfo {volume} {427}},\ \bibinfo {pages} {615} (\bibinfo {year} {2004})}\BibitemShut {NoStop}%
\bibitem [{\citenamefont {Xu}\ \emph {et~al.}(2007)\citenamefont {Xu}, \citenamefont {Manipatruni}, \citenamefont {Schmidt}, \citenamefont {Shakya},\ and\ \citenamefont {Lipson}}]{xu200712}%
  \BibitemOpen
  \bibfield  {author} {\bibinfo {author} {\bibfnamefont {Q.}~\bibnamefont {Xu}}, \bibinfo {author} {\bibfnamefont {S.}~\bibnamefont {Manipatruni}}, \bibinfo {author} {\bibfnamefont {B.}~\bibnamefont {Schmidt}}, \bibinfo {author} {\bibfnamefont {J.}~\bibnamefont {Shakya}},\ and\ \bibinfo {author} {\bibfnamefont {M.}~\bibnamefont {Lipson}},\ }\bibfield  {title} {\bibinfo {title} {12.5 gbit/s carrier-injection-based silicon micro-ring silicon modulators},\ }\href@noop {} {\bibfield  {journal} {\bibinfo  {journal} {Optics express}\ }\textbf {\bibinfo {volume} {15}},\ \bibinfo {pages} {430} (\bibinfo {year} {2007})}\BibitemShut {NoStop}%
\bibitem [{\citenamefont {Manipatruni}\ \emph {et~al.}(2007)\citenamefont {Manipatruni}, \citenamefont {Xu}, \citenamefont {Schmidt}, \citenamefont {Shakya},\ and\ \citenamefont {Lipson}}]{manipatruni2007high}%
  \BibitemOpen
  \bibfield  {author} {\bibinfo {author} {\bibfnamefont {S.}~\bibnamefont {Manipatruni}}, \bibinfo {author} {\bibfnamefont {Q.}~\bibnamefont {Xu}}, \bibinfo {author} {\bibfnamefont {B.}~\bibnamefont {Schmidt}}, \bibinfo {author} {\bibfnamefont {J.}~\bibnamefont {Shakya}},\ and\ \bibinfo {author} {\bibfnamefont {M.}~\bibnamefont {Lipson}},\ }\bibfield  {title} {\bibinfo {title} {High speed carrier injection 18 gb/s silicon micro-ring electro-optic modulator},\ }in\ \href@noop {} {\emph {\bibinfo {booktitle} {LEOS 2007-IEEE Lasers and Electro-Optics Society Annual Meeting Conference Proceedings}}}\ (\bibinfo {organization} {IEEE},\ \bibinfo {year} {2007})\ pp.\ \bibinfo {pages} {537--538}\BibitemShut {NoStop}%
\bibitem [{\citenamefont {Manolatou}\ and\ \citenamefont {Lipson}(2006)}]{manolatou2006all}%
  \BibitemOpen
  \bibfield  {author} {\bibinfo {author} {\bibfnamefont {C.}~\bibnamefont {Manolatou}}\ and\ \bibinfo {author} {\bibfnamefont {M.}~\bibnamefont {Lipson}},\ }\bibfield  {title} {\bibinfo {title} {All-optical silicon modulators based on carrier injection by two-photon absorption},\ }\href@noop {} {\bibfield  {journal} {\bibinfo  {journal} {Journal of lightwave technology}\ }\textbf {\bibinfo {volume} {24}},\ \bibinfo {pages} {1433} (\bibinfo {year} {2006})}\BibitemShut {NoStop}%
\bibitem [{\citenamefont {Zhang}\ \emph {et~al.}(2020)\citenamefont {Zhang}, \citenamefont {Ebert}, \citenamefont {Chen}, \citenamefont {Reynolds}, \citenamefont {Yan}, \citenamefont {Du}, \citenamefont {Banakar}, \citenamefont {Tran}, \citenamefont {Debnath}, \citenamefont {Littlejohns} \emph {et~al.}}]{zhang2020integration}%
  \BibitemOpen
  \bibfield  {author} {\bibinfo {author} {\bibfnamefont {W.}~\bibnamefont {Zhang}}, \bibinfo {author} {\bibfnamefont {M.}~\bibnamefont {Ebert}}, \bibinfo {author} {\bibfnamefont {B.}~\bibnamefont {Chen}}, \bibinfo {author} {\bibfnamefont {J.~D.}\ \bibnamefont {Reynolds}}, \bibinfo {author} {\bibfnamefont {X.}~\bibnamefont {Yan}}, \bibinfo {author} {\bibfnamefont {H.}~\bibnamefont {Du}}, \bibinfo {author} {\bibfnamefont {M.}~\bibnamefont {Banakar}}, \bibinfo {author} {\bibfnamefont {D.}~\bibnamefont {Tran}}, \bibinfo {author} {\bibfnamefont {K.}~\bibnamefont {Debnath}}, \bibinfo {author} {\bibfnamefont {C.}~\bibnamefont {Littlejohns}}, \emph {et~al.},\ }\bibfield  {title} {\bibinfo {title} {Integration of low loss vertical slot waveguides on soi photonic platforms for high efficiency carrier accumulation modulators},\ }\href@noop {} {\bibfield  {journal} {\bibinfo  {journal} {Optics Express}\ }\textbf {\bibinfo {volume} {28}},\ \bibinfo {pages} {23143} (\bibinfo {year} {2020})}\BibitemShut {NoStop}%
\bibitem [{\citenamefont {Zhang}\ \emph {et~al.}(2023)\citenamefont {Zhang}, \citenamefont {Ebert}, \citenamefont {Li}, \citenamefont {Chen}, \citenamefont {Yan}, \citenamefont {Du}, \citenamefont {Banakar}, \citenamefont {Tran}, \citenamefont {Littlejohns}, \citenamefont {Scofield} \emph {et~al.}}]{zhang2023harnessing}%
  \BibitemOpen
  \bibfield  {author} {\bibinfo {author} {\bibfnamefont {W.}~\bibnamefont {Zhang}}, \bibinfo {author} {\bibfnamefont {M.}~\bibnamefont {Ebert}}, \bibinfo {author} {\bibfnamefont {K.}~\bibnamefont {Li}}, \bibinfo {author} {\bibfnamefont {B.}~\bibnamefont {Chen}}, \bibinfo {author} {\bibfnamefont {X.}~\bibnamefont {Yan}}, \bibinfo {author} {\bibfnamefont {H.}~\bibnamefont {Du}}, \bibinfo {author} {\bibfnamefont {M.}~\bibnamefont {Banakar}}, \bibinfo {author} {\bibfnamefont {D.~T.}\ \bibnamefont {Tran}}, \bibinfo {author} {\bibfnamefont {C.~G.}\ \bibnamefont {Littlejohns}}, \bibinfo {author} {\bibfnamefont {A.}~\bibnamefont {Scofield}}, \emph {et~al.},\ }\bibfield  {title} {\bibinfo {title} {Harnessing plasma absorption in silicon mos ring modulators},\ }\href@noop {} {\bibfield  {journal} {\bibinfo  {journal} {Nature Photonics}\ }\textbf {\bibinfo {volume} {17}},\ \bibinfo {pages} {273} (\bibinfo {year} {2023})}\BibitemShut {NoStop}%
\bibitem [{\citenamefont {Gevorgyan}\ \emph {et~al.}(2021)\citenamefont {Gevorgyan}, \citenamefont {Khilo}, \citenamefont {Wade}, \citenamefont {Stojanovi{\'c}},\ and\ \citenamefont {Popovi{\'c}}}]{gevorgyan2021moscap}%
  \BibitemOpen
  \bibfield  {author} {\bibinfo {author} {\bibfnamefont {H.}~\bibnamefont {Gevorgyan}}, \bibinfo {author} {\bibfnamefont {A.}~\bibnamefont {Khilo}}, \bibinfo {author} {\bibfnamefont {M.~T.}\ \bibnamefont {Wade}}, \bibinfo {author} {\bibfnamefont {V.~M.}\ \bibnamefont {Stojanovi{\'c}}},\ and\ \bibinfo {author} {\bibfnamefont {M.~A.}\ \bibnamefont {Popovi{\'c}}},\ }\bibfield  {title} {\bibinfo {title} {Moscap ring modulator with 1.5 $\mu$m radius, 8.5 thz fsr and 30 ghz/v shift efficiency in a 45 nm soi cmos process},\ }in\ \href@noop {} {\emph {\bibinfo {booktitle} {2021 Optical Fiber Communications Conference and Exhibition (OFC)}}}\ (\bibinfo {organization} {IEEE},\ \bibinfo {year} {2021})\ pp.\ \bibinfo {pages} {1--3}\BibitemShut {NoStop}%
\bibitem [{\citenamefont {Luo}\ \emph {et~al.}(2011)\citenamefont {Luo}, \citenamefont {Wiederhecker}, \citenamefont {Cardenas}, \citenamefont {Poitras},\ and\ \citenamefont {Lipson}}]{Luo:11}%
  \BibitemOpen
  \bibfield  {author} {\bibinfo {author} {\bibfnamefont {L.-W.}\ \bibnamefont {Luo}}, \bibinfo {author} {\bibfnamefont {G.~S.}\ \bibnamefont {Wiederhecker}}, \bibinfo {author} {\bibfnamefont {J.}~\bibnamefont {Cardenas}}, \bibinfo {author} {\bibfnamefont {C.}~\bibnamefont {Poitras}},\ and\ \bibinfo {author} {\bibfnamefont {M.}~\bibnamefont {Lipson}},\ }\bibfield  {title} {\bibinfo {title} {High quality factor etchless silicon photonic ring resonators},\ }\href {https://doi.org/10.1364/OE.19.006284} {\bibfield  {journal} {\bibinfo  {journal} {Opt. Express}\ }\textbf {\bibinfo {volume} {19}},\ \bibinfo {pages} {6284} (\bibinfo {year} {2011})}\BibitemShut {NoStop}%
\bibitem [{\citenamefont {Belogolovskii}\ \emph {et~al.}(2025)\citenamefont {Belogolovskii}, \citenamefont {Rahman}, \citenamefont {Johnson}, \citenamefont {Fedorov}, \citenamefont {Grieco}, \citenamefont {Alic}, \citenamefont {Ndao}, \citenamefont {Yu},\ and\ \citenamefont {Fainman}}]{belogolovskii2025large}%
  \BibitemOpen
  \bibfield  {author} {\bibinfo {author} {\bibfnamefont {D.}~\bibnamefont {Belogolovskii}}, \bibinfo {author} {\bibfnamefont {M.~M.}\ \bibnamefont {Rahman}}, \bibinfo {author} {\bibfnamefont {K.}~\bibnamefont {Johnson}}, \bibinfo {author} {\bibfnamefont {V.}~\bibnamefont {Fedorov}}, \bibinfo {author} {\bibfnamefont {A.}~\bibnamefont {Grieco}}, \bibinfo {author} {\bibfnamefont {N.}~\bibnamefont {Alic}}, \bibinfo {author} {\bibfnamefont {A.}~\bibnamefont {Ndao}}, \bibinfo {author} {\bibfnamefont {P.~K.}\ \bibnamefont {Yu}},\ and\ \bibinfo {author} {\bibfnamefont {Y.}~\bibnamefont {Fainman}},\ }\bibfield  {title} {\bibinfo {title} {Large bidirectional refractive index change in silicon-rich nitride via visible light trimming},\ }\href@noop {} {\bibfield  {journal} {\bibinfo  {journal} {Advanced Optical Materials}\ }\textbf {\bibinfo {volume} {13}},\ \bibinfo {pages} {2403420} (\bibinfo {year} {2025})}\BibitemShut {NoStop}%
\bibitem [{\citenamefont {Xie}\ \emph {et~al.}(2021)\citenamefont {Xie}, \citenamefont {Frankis}, \citenamefont {Bradley},\ and\ \citenamefont {Knights}}]{Xie:21}%
  \BibitemOpen
  \bibfield  {author} {\bibinfo {author} {\bibfnamefont {Y.}~\bibnamefont {Xie}}, \bibinfo {author} {\bibfnamefont {H.~C.}\ \bibnamefont {Frankis}}, \bibinfo {author} {\bibfnamefont {J.~D.~B.}\ \bibnamefont {Bradley}},\ and\ \bibinfo {author} {\bibfnamefont {A.~P.}\ \bibnamefont {Knights}},\ }\bibfield  {title} {\bibinfo {title} {Post-fabrication resonance trimming of si3n4 photonic circuits via localized thermal annealing of a sputter-deposited sio2 cladding},\ }\href {https://doi.org/10.1364/OME.426775} {\bibfield  {journal} {\bibinfo  {journal} {Opt. Mater. Express}\ }\textbf {\bibinfo {volume} {11}},\ \bibinfo {pages} {2401} (\bibinfo {year} {2021})}\BibitemShut {NoStop}%
\end{thebibliography}%

\clearpage
\onecolumngrid 
\begin{center}
  {\large\bfseries Supplementary Material for}\\[1.2ex]
  {\large\bfseries Demonstration and Non-volatile Trimming of a Highly Parallel,\\
  High Capacity Silicon Micro-disk Transmitter}
\end{center}

\vspace{1.2em}

\begin{center}
  Chao Luan$^{1,*}$, Alex Sludds$^{1}$, Chao Li$^{1}$, Ian Christen$^{1}$,
  Ryan Hamerly$^{1,2}$, Dirk Englund$^{1}$\\[0.7em]
  \textit{$^{1}$Research Laboratory of Electronics, MIT, Cambridge, MA 02139, USA}\\
  \textit{$^{2}$NTT Research Inc., PHI Laboratories, 940 Stewart Drive, Sunnyvale, CA 94085, USA}
\end{center}

\section{Vertical p-n junction microdisk modulator}
\label{sec:s1}
Using the free-carrier dispersion effect, a variety of silicon-based resonant EOIMs have been demonstrated~\cite{rahim2021taking, reed2010silicon, xu2005micrometre, yuan20245, timurdogan2014ultralow, witzens2018high, liu2004high}. The carrier-injection-based modulators offer significant amplitude modulation efficiency but are inherently limited in bandwidth due to the low mobility speed of the injected carriers~\cite{xu200712, manipatruni2007high, manolatou2006all}. The carrier accumulation modulators, employing metal-oxide-semiconductor (MOS) capacitors, provide efficient modulation and high bandwidth; however, they require complex fabrication processes involving additional thin oxide layers and poly-silicon depositions~\cite{zhang2020integration, zhang2023harnessing, gevorgyan2021moscap}. Conversely, carrier depletion resonant EOIMs operating under reverse-biased voltages achieve low energy consumption and high bandwidth but typically suffer from lower modulation efficiency. To address these limitations, vertical-type p–n junction microdisk EOIMs have been developed. The vertical junction configuration enables internal \(n^{+}\) and \(p^{+}\) electrical contacts, eliminating the necessity for a ridge waveguide structure and thus retaining a smooth, hard outer wall for the silicon microdisk resonator. This hard outer wall significantly enhances the optical mode confinement, allowing for the realization of extremely compact microdisk modulators with micrometer-scale diameters. Consequently, this design greatly reduces device capacitance, enabling large bandwidth and low energy consumption modulation. Additionally, the vertical junction microdisk modulators provide substantial spatial overlap between carriers and the optical mode, resulting in enhanced modulation efficiency~\cite{daudlin2025three, yuan20245, timurdogan2014ultralow, rizzo2022petabit, rizzo2023massively}.

\section{Characterization of the micro-disk modulator}

\subsection{Subsection 1: ER, IL, and OMA}
\label{sec:s21}

\begin{figure}[!htbp]
  \centering
  \includegraphics[width=0.45\textwidth]{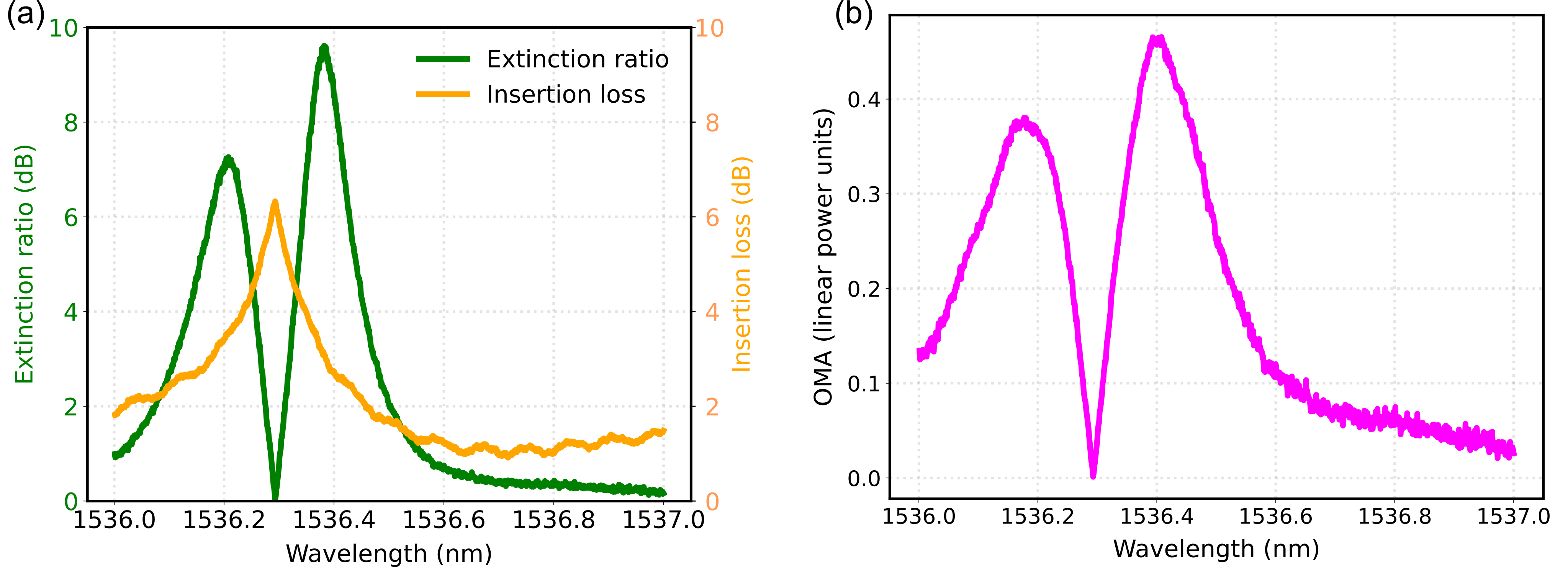}
  \caption{ER, IL and OMA results of the microdisk transmitter under a 2-V swing voltage.}
  \label{fig:sfig21}
\end{figure}

Figure~\ref{fig:sfig21} summarizes the static optical-power metrics of the 64-channel microdisk transmitters when driven with a 2 V\(_\text{pp}\) electrical swing. For each wavelength we extract

\begin{itemize}
    \item the IL: the on-resonance transmission of the ‘1’ level relative to the input fiber power;
    \item the ER: \(\mathrm{ER}=P_{\!1}/P_{\!0}\) expressed in decibels, where \(P_{\!1}\) and \(P_{\!0}\) are the optical powers in the logical ‘1’ and ‘0’ states, respectively; and
    \item the OMA, defined in linear units as \(\mathrm{OMA}=P_{\!1}-P_{\!0}\).
\end{itemize}

The devices exhibit a largest ER of \(9.8\ \text{dB}\) and an IL of \(2.6\ \text{dB}\). The corresponding average OMA is -3.2 dB response to the input power in a 2-V voltage swing.

\subsection{Extracting the waveguide absorption coefficient and refractive-index change}
\label{sec:s22}
\begin{figure*}[!htbp]
\centering
\includegraphics[width=0.4\textwidth]{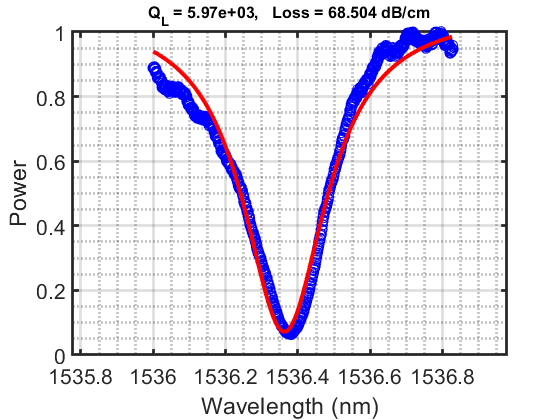}
\caption{Measured and fitted transmission of the microdisk, showing the absorption coefficient and quality factor of the microdisk under the rever biased voltage at -1 V.}
\label{fig:7}
\end{figure*}

The through-port transmission of an all-pass micro-ring resonator is
\begin{equation}
T(\phi)=\frac{a^{2}-2ar\cos\phi+r^{2}} {1-2ar\cos\phi+(ar)^{2}},
\label{eq:Tphi}
\end{equation}
with $a=\exp(-\alpha L/2)$ and $L=2\pi R$.
Near an isolated resonance $\lambda=\lambda_{0}$ we expand $\phi\simeq\phi_{0}+2\pi L(\lambda-\lambda_{0})/\!\bigl(\lambda_{0}^{2}/n_{\mathrm{g}}\bigr)$;
Eq.~\eqref{eq:Tphi} then reduces to the Lorentzian form
\begin{equation}
T(\lambda)=1-\frac{A}{1+\bigl[2(\lambda-\lambda_{0})/\Delta\lambda\bigr]^{2}},
\label{eq:Lorentz}
\end{equation}
whose three fit parameters provide the \textbf{loaded quality factor} $Q_{\mathrm{L}}=\lambda_{0}/\Delta\lambda$ and the dip depth $T_{\min}=1-A$.
\vspace{2em}
\paragraph{Intrinsic \(Q\):}
Following~~\cite{Luo:11}, the intrinsic quality factor is
\begin{equation}
Q_{\mathrm{int}} =\frac{2\,Q_{\mathrm{L}}}{1+\sqrt{T_{\min}}},
\label{eq:Qint}
\end{equation}
\vspace{2em}
\paragraph{Propagation loss:}
The round-trip intrinsic loss then yields the linear wave-guide attenuation
\begin{equation}
\boxed{%
\alpha =\frac{2\pi n_{\mathrm{g}}}{Q_{\mathrm{int}}\lambda_{0}},
\qquad
\alpha_{\mathrm{dB/cm}}=4.343\times10^{2}\,\alpha
}
\label{eq:alpha}
\end{equation}
Applying Eqs.~\eqref{eq:Tphi}–\eqref{eq:alpha} to all the reverse-biased spectra and subtracting gives the bias-induced absorption
change \(\Delta\alpha=\alpha_{\text{biased${}_{1}$}}-\alpha_{\text{biased${}_{2}$}}\).
\vspace{2em}
\paragraph{Effective–index change:}
The resonance condition for an $m$-th order mode is $m\lambda_0=n_{\mathrm{eff}}L$.
Differentiating and holding $L$ fixed gives $\Delta n_{\mathrm{eff}}/n_{\mathrm{eff}}=\Delta\lambda/\lambda_0$,
so the bias-induced effective-index change is
\begin{equation}
\boxed{\;
\Delta n_{\mathrm{eff}}
      = n_{\mathrm{g}}\,
        \frac{\Delta\lambda}{\lambda_{0}}
\;}
\label{eq:dn}
\end{equation}
where $\Delta\lambda=\lambda_{\text{biased1}}-\lambda_{\text{biased2}}$ is the measured
wavelength shift and $n_{\mathrm{g}}$ is the group index obtained from the waveguide dispersion calculation.  
Combining Eqs.~\eqref{eq:alpha} and~\eqref{eq:dn} therefore yields both the absorption and refractive-index changes of the micro-disk from the same set of transmission spectra.
Fig.~\ref{fig:7} shows an example of the measured (blue) and fitted (red) transmission spectrum of the microdisk modulator at revised bias voltage of -1 V, which returns an absorption coefficient of 6.85 dB/mm.

\subsection{Electrical response (\texorpdfstring{$S_{11}$}{S11}) of the lumped micro-disk and circuit parameter extraction}
\label{sec:s23}
\begin{figure*}[!htbp]
  \centering
  \includegraphics[width=0.85\textwidth]{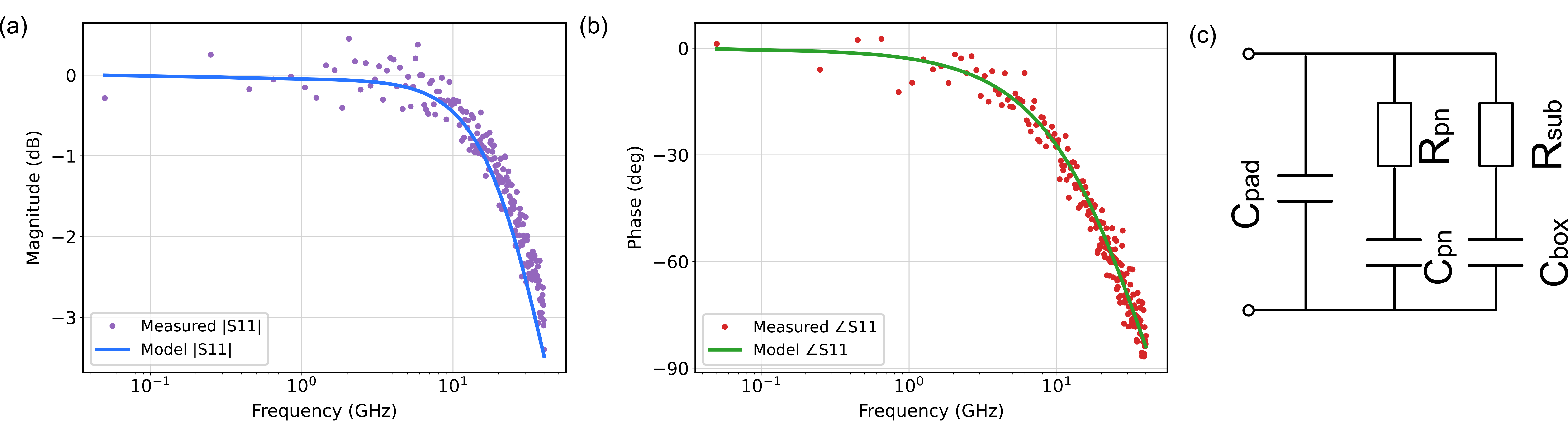}
  \caption{Measured (dots) and fitted (solid line) $S_{11}$ magnitude of the micro-disk.}
  \label{fig:sfig23}
\end{figure*}

Figure~\ref{fig:sfig23} shows the three-branch pad model that accurately fits the small-signal RF behavior of the micro-disk modulator:
\begin{equation}
\begin{aligned}
Y_{\mathrm{pad}}(\omega)  &=  j\omega C_{\mathrm{pad}},\\[4pt]
Y_{\mathrm{pn}}(\omega)   &= \dfrac{1}{\,R_{\mathrm{pn}} + \dfrac{1}{j\omega C_{\mathrm{pn}}}},\\[6pt]
Y_{\mathrm{sub}}(\omega)  &= \dfrac{1}{\,R_{\mathrm{sub}} + \dfrac{1}{j\omega C_{\mathrm{box}}}}.
\end{aligned}
\label{eq:Ybranches}
\end{equation}
The total shunt admittance is
\begin{equation}
Y_{\mathrm{tot}}(\omega)=Y_{\mathrm{pad}}+Y_{\mathrm{pn}}+Y_{\mathrm{sub}},
\label{eq:Ytot}
\end{equation}
the input impedance is $Z_{\mathrm{in}}=1/Y_{\mathrm{tot}}$, and with a $50\;\Omega$ reference
\begin{equation}
S_{11}(\omega)=\dfrac{Z_{\mathrm{in}}(\omega)-Z_{0}}{Z_{\mathrm{in}}(\omega)+Z_{0}},
\qquad Z_{0}=50~\Omega.
\label{eq:S11model}
\end{equation}
A calibrated VNA measures the complex $S_{11}^{\text{meas}}(\omega)$ from \(50~\mathrm{MHz}\) to \(40.5~\mathrm{GHz}\) using a GS probe. The substrate contact is left floating so that the entire pad model appears between signal and ground. Although all five parameters can be obtained from a single non-linear fit, we follow the procedure of~\cite{rizzo2023massively,daudlin2025three}: the pad resistance is fixed to the software simulation value while the remaining parameters are fitted from the measurements. Figure~\ref{fig:sfig23} compares the measured and fitted $S_{11}$; the resulting lumped parameters are summarized in Table~\ref{tab:s11}.

\begin{table}[h]
  \centering
  \caption{Extracted parameters of the microdisk circuit.}
  \vspace{1.2em}  
  \label{tab:s11}
  \setlength{\tabcolsep}{14pt}
  \renewcommand{\arraystretch}{1.25}
  \begin{tabular}{|c|c|c|c|c|}
    \hline
    \textbf{$C_{\mathrm{pn}}$} & \textbf{$R_{\mathrm{pn}}$} &
    \textbf{$C_{\mathrm{box}}$} & \textbf{$R_{\mathrm{sub}}$} &
    \textbf{$C_{pad}$} \\ \hline
    75\,fF & 23\,$\Omega$ & 34\,fF & 19\,k$\Omega$ & 3.5\,fF \\ \hline
  \end{tabular}
\end{table}

From the above circuit parameter values, the RC determined bandwidth S\(_2\)\(_1\) was calculated, and the final bandwidth, which is determined by the RC bandwidth and the photon lifetime, was calculated and shown in the main paper. It was noted that the final bandwidth is also determined by the input laser wavelength, a smaller wavelength-resonance difference will decrease the final measured bandwidth.






\section{Non-volatile trimming}
\subsection{Subsection 1: Laser annealing }
\label{sec:s41}
\begin{figure}[!htbp]
  \centering
  \includegraphics[width=0.5\textwidth]{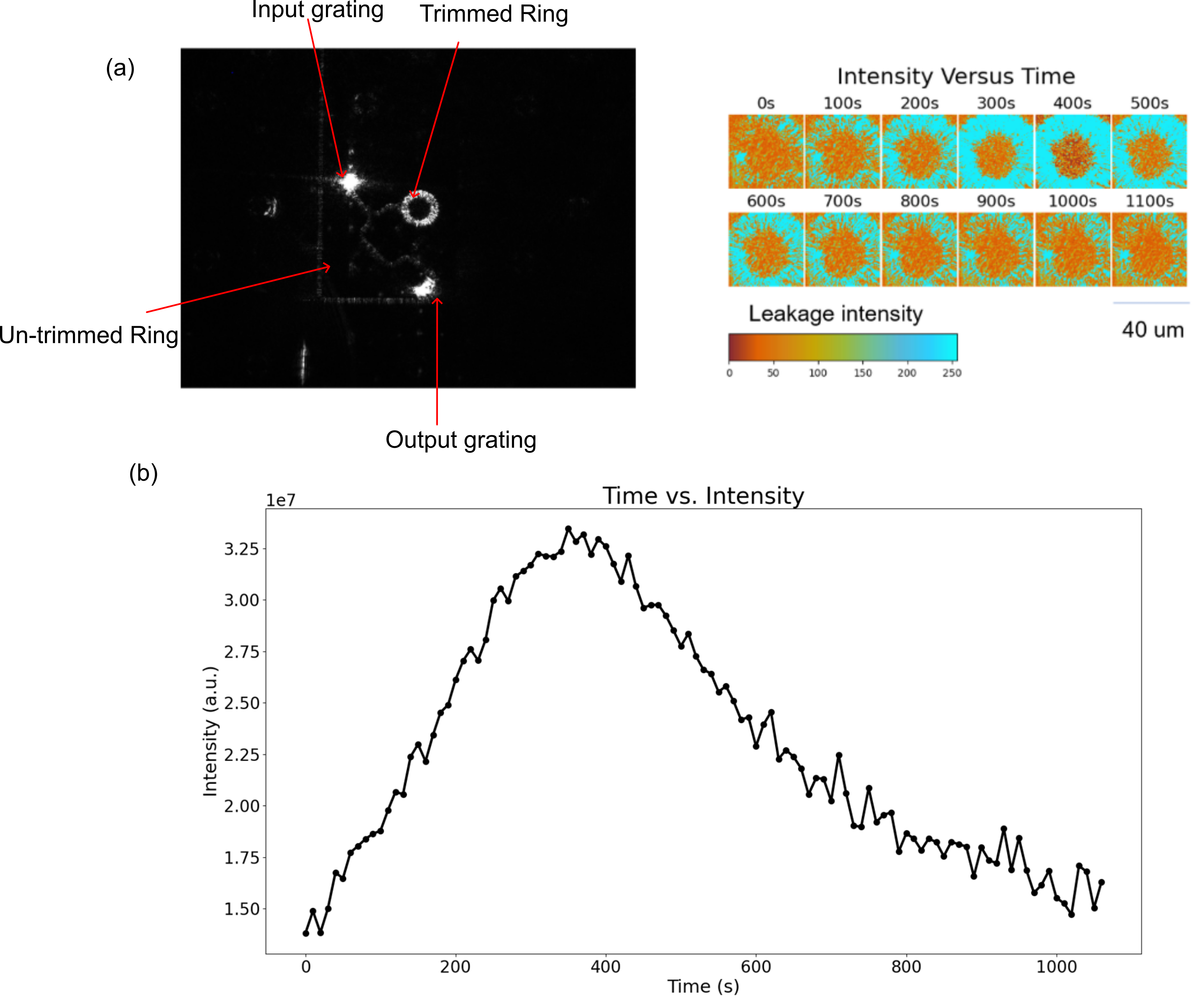}
  \caption{Laser annealing validation in cladding covered SiN E/O modulator.}
  \label{fig:sfig41}
\end{figure}

The laser annealing was first verified in a SiN microring E/O modulator, the microring is covered with top thick cladding, through the laser annealing, the cladding bonding angle changes, this will increase the cladding compaction, and increase the refractive index. During the annealing process, we capture the leakage intensity from above the microring directly, The measured leakage intensity versus time is shown in Figure~\ref{fig:sfig41}, as clearly shown, the microring started with off resonance, through the trimming, the microring gradually becomes to in resonance state, and we can see a clear 'ring shape' leakage, further trimming the microring will make it becomes off resonance again, the leakage intensity gradually increase, when trimmed to the in resonance state, the leakage intensity reach to the highest value, and then gradually decrease with the trimming laser on.

\begin{figure}[!htbp]
  \centering
  \includegraphics[width=0.3\textwidth]{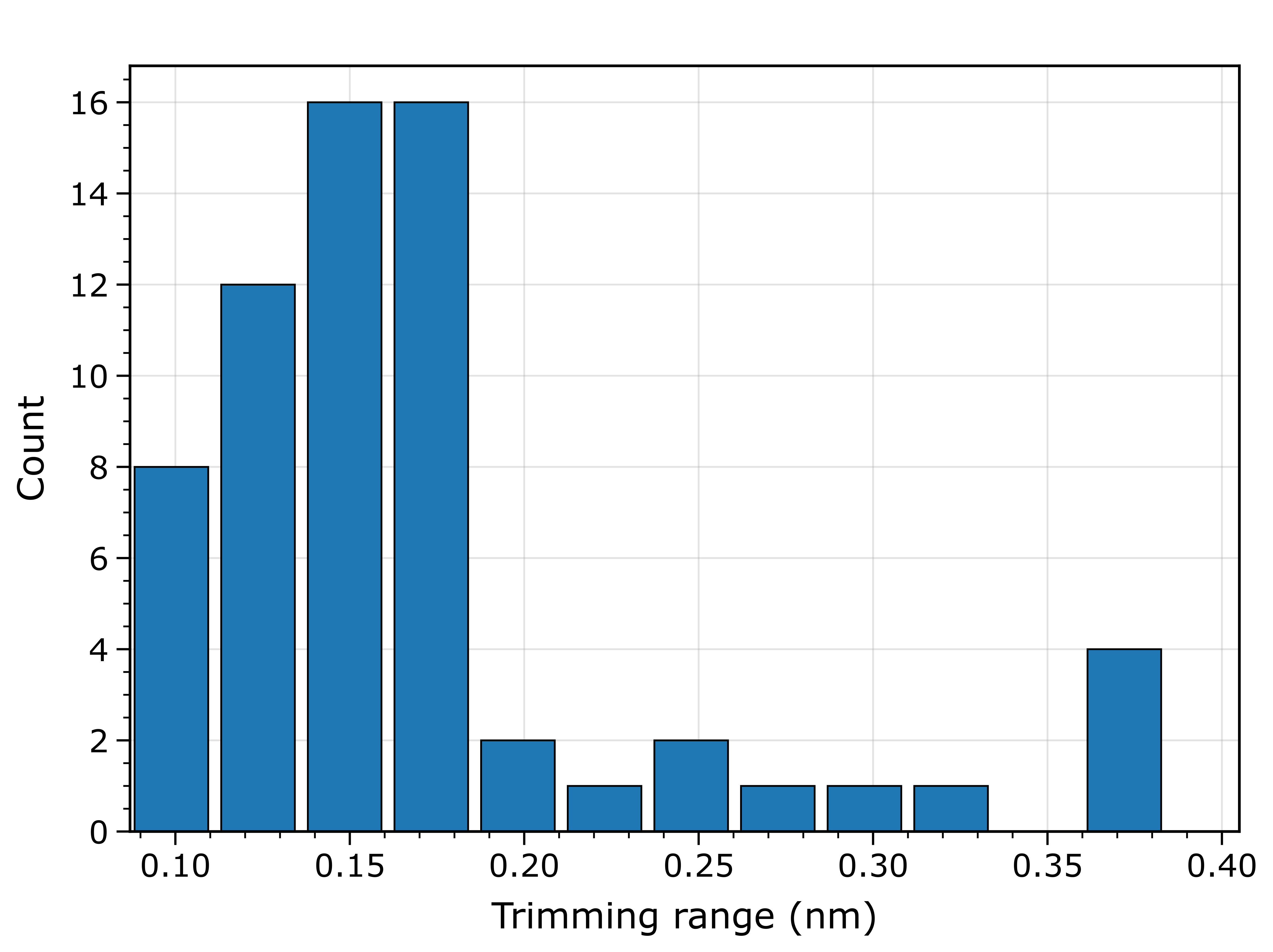}
  \caption{Laser annealing results among different channels.}
  \label{fig:sfig43}
\end{figure}

In the silicon micro-disk the out-of-plane scatter port is masked by the integrated heater, so the resonance shift is monitored at the grating-coupler output.
A laser spot with a $1~\upmu\text{m}$ radius ($\approx 3~\upmu\text{m}^2$) is used for trimming; its power density is $\sim\!1~\text{mW}\,\upmu\text{m}^{-2}$, well above the SiO$_2$ densification threshold.
Because the optical mode in silicon is much more tightly confined than in SiN, its evanescent overlap with the cladding—and hence the index change produced by compaction—is smaller. Consequently, the cavity resonance shift is between $\sim\!100~\text{pm}$ to $\sim\!380~\text{pm}$, changes among different microdisk due to the SiO$_2$ quality different of among different devices.  

\newpage
\subsection{Subsection 2: Reduce the thermal tunning energy consumption}
\label{sec:s42}
The thermal energy required to tune a resonator scales with the wavelength shift needed to reach the target,
\(\Delta\lambda = \lambda_{\mathrm{tar}} - \lambda\).
Because our trimming method provides only a red shift, devices whose initial resonance exceeds the target
\((\lambda_i > \lambda_{\mathrm{tar}})\) are excluded from analysis.

For microdisks with \(\lambda_i < \lambda_{\mathrm{tar}}\), we apply a trimming cap \(C\) 
and define the post-trim resonance for device \(i\) as
\begin{equation}
  \lambda_i'(C) = \min\!\bigl(\lambda_i + C,\ \lambda_{\mathrm{tar}}\bigr).
  \label{eq:posttrim}
\end{equation}
We then compute the ensemble-average resonance over all included devices,
\begin{equation}
  \bar{\lambda}(C) = \langle \lambda_i'(C) \rangle,
  \label{eq:avg}
\end{equation}
and the residual detuning relative to the target,
\begin{equation}
  \varepsilon(C) = \lambda_{\mathrm{tar}} - \bar{\lambda}(C).
  \label{eq:residual}
\end{equation}
Since the thermal energy required after trimming is proportional to the remaining wavelength error, the relative energy saving is,
\begin{equation}
  S(C) = 1 - \frac{\varepsilon(C)}{\varepsilon(0)},
  \label{eq:saving}
\end{equation}
where \(\varepsilon(0)\) is the untrimmed residual detuning computed from the pre-trim ensemble average.

\subsection{Subsection 3: Reduce the design redundancy}
\label{sec:s43}

We model each of the \(K\) WDM channels as provisioned with redundancy \(r\) (i.e., up to \(r\) rings per channel).
Assuming each ring independently lands within the target passband with probability \(p\),
the channel success probability is
\begin{equation}
  p_{\mathrm{ch}}(p,r) = 1 - (1-p)^{r}.
  \label{eq:p_channel}
\end{equation}
Hence, the expected number of usable WDM channels is
\begin{equation}
  \mathbb{E}[N] = K \, p_{\mathrm{ch}}(p,r).
  \label{eq:expected_channels}
\end{equation}

For a target of at least \(N_{\mathrm{tgt}}\) usable channels in expectation, the minimal redundancy that satisfies
\(\mathbb{E}[N] \ge N_{\mathrm{tgt}}\) is
\begin{equation}
  r^\star(p; N_{\mathrm{tgt}},K) \;=\;
  \left\lceil \frac{\ln\!\bigl(1 - N_{\mathrm{tgt}}/K\bigr)}{\ln(1-p)} \right\rceil.
  \label{eq:r_star}
\end{equation}

Trimming increases the per-ring success probability from \(p_{\mathrm{pre}}=0.0468\) to \(p_{\mathrm{post}}=0.14\).
Using \eqref{eq:r_star} with different \(K\) number, the required redundancy was calculated, as indicated in Figure~\ref{fig:sfig41}.

\newpage
\subsection{Subsection 4: Thermal annealing}
\label{sec:s42}
Thermal annealing is another widely used post-fabrication trimming technique for resonant silicon photonic devices~\cite{chen2025thermally,wu2025lossless}.
By holding the chip at several hundred degree, the SiO$_2$ cladding densifies and its refractive index increases, permanently shifting the cavity resonance.
Furnace exposures of $\sim10\text{–}15~\mathrm{min}$ at $400\text{–}500^{\circ}\mathrm{C}$ have been shown to tune passive SiN microrings by more than one free-spectral range with negligible excess loss~\cite{chen2025thermally}.

\medskip
\noindent\textbf{Limitation—high temperature.}
The high process temperature that makes annealing effective for passive structures also limits its applicability to \emph{active} components.
Above $\sim300^{\circ}\mathrm{C}$ aluminium or Titanium heaters can reflow, dopant profiles in p–n junctions smear out, silicide contacts degrade, metal–oxide capacitors, and \(p-n\) diode suffer irreversible damage.
Consequently furnace annealing is unsuitable for modulators, photodetectors, or devices that incorporate heterogeneous III–V stacks, where even brief high-temperature exposure can compromise speed or reliability.
These constraints motivate low-temperature, highly localized alternatives—such as the laser-induced cladding compaction demonstrated in this work—which trim individual resonators without subjecting the rest of the photonic circuit to damaging thermal budgets.
We also implement the thermal annealing method in the current chip, we set the thermal heater driving voltage to 8-V, the calculated temperature in the silicon waveguide is about $300 ^{\circ}\mathrm{C}$, turning off the heater driving voltage, we find a 200 pm wavelength red shift. To avoid the potential damage to the modulator, we didn't use this thermal annealing method during the experiment. 
The issue of the thermal trimmed wavelength shifting is the long time stability, we monitor the resonant wavelength of the thermal trimmed device, and find the resonant wavelength gradually returned to the initial state after 2 weeks, which has been found in other journals~\cite{belogolovskii2025large, Xie:21}.

\subsection{Subsection 5: Silicon Oxidation}
\label{sec:s43}
\begin{figure}[!htbp]
    \centering
    \includegraphics[width=0.3\textwidth]{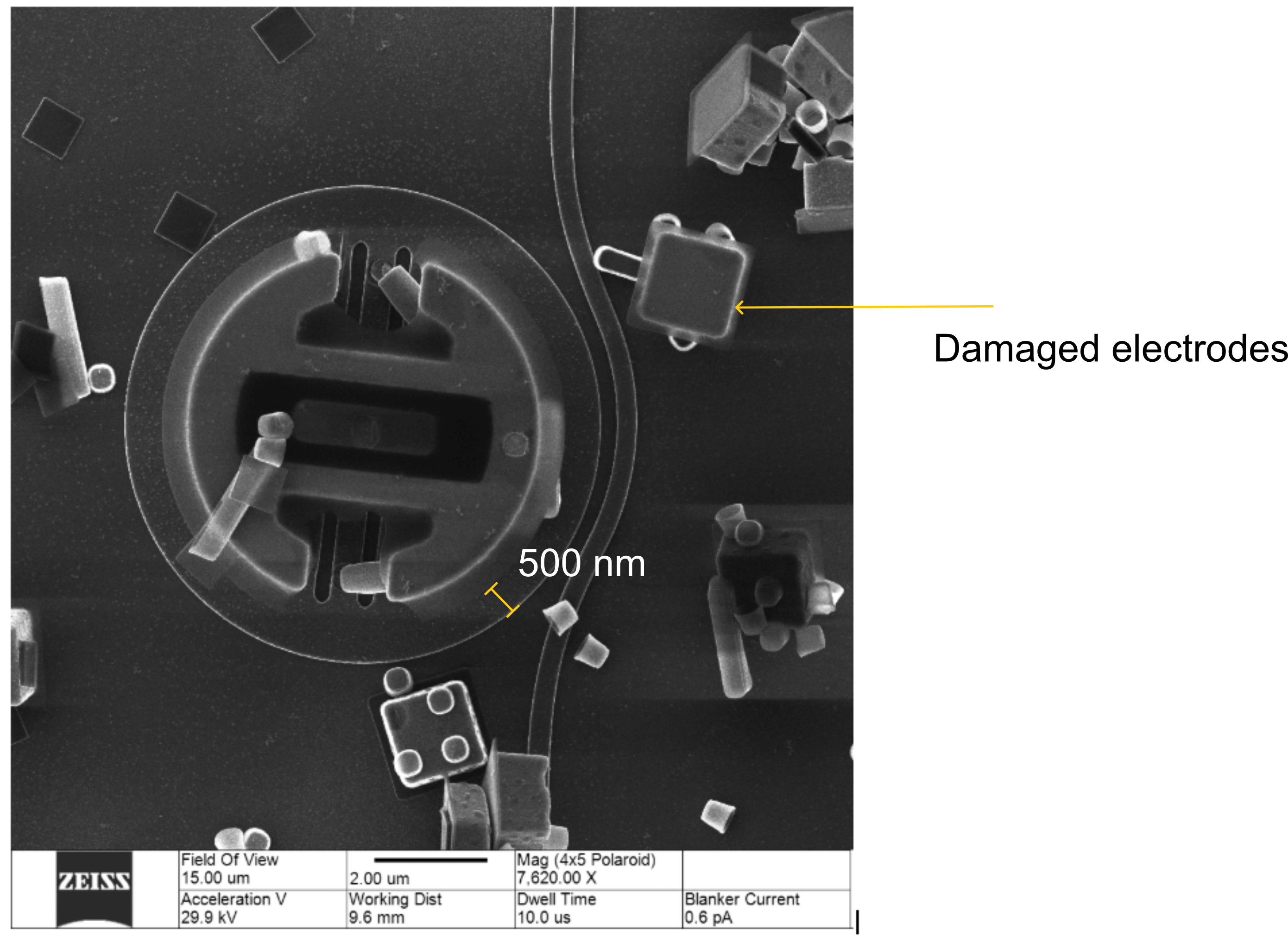}
    \caption{SEM image of the wet etched microdisk, after the resist removing, the electrcode was also damaged.}
    \label{fig:sfig44}
\end{figure}

Oxidizing the silicon core is an attractive post-fabrication technique because the accompanying reduction of the effective refractive index produces a permanent blue-shift of the resonance. In our devices, however, the microrings are buried beneath a thick upper SiO$_2$ cladding and overlaid by a metal heater wires and contact. We therefore explored a two-step process—wet removal of the cladding followed by thermal oxidation of the exposed waveguide.
Before the wet etching we open a window to protect the other part of the device. The window was opened using the lithography and developing. Due to the charging effect, the window was opened using the UV maskless lithography. However, even though the other part of the device is protected using the resist, the HF-based etchlant that stripped the oxide also attacked the heater electrodes, leaving the device electrically open and severely degrading the modulator (Figure~\ref{fig:sfig44}). Owing to this collateral damage we abandoned the oxide-removal/oxidation route.

\newpage
\subsection{Subsection 4: PCM deposition}
\label{sec:s7}
\begin{figure}[!htbp]
  \centering
  \includegraphics[width=0.8\textwidth]{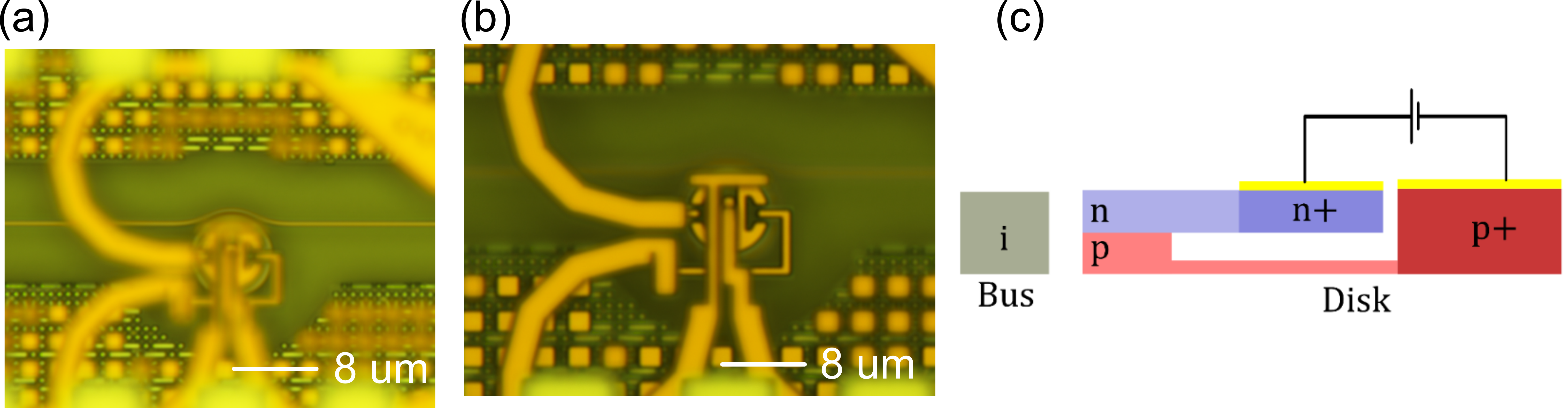}
  \caption{Microscope image and schematic of the microdisk transmitter, where the silicon exposure region is too small and the PCM couldn't be deposited.}
  \label{fig:sfig45}
\end{figure}

Recent years have witnessed significant interest in low-loss phase-change materials (PCMs) such as Sb$_2$Se$_3$ (SbSe) and Sb$_2$S$_3$ for photonic applications. Their non-volatile nature---maintaining their state without a continuous voltage supply---further enhances their suitability for the trimming application in integrated photonic devices.
We also tried to investigate the trimming effect in the current compact microdisk EOIM, however, several facts limits this application:
\begin{enumerate}
  \item \textbf{Metal coverage and limited exposure area.} Unlike other types microring, the microdisk radius is small and most of its surface is already occupied by the p$^+$/n$^+$ metal contacts(Figure~\ref{fig:sfig45}(a,b)). This leaves too little exposed Si (which is 500 nm in our case, as shown in the SEM image) area to open windows, deposit, and pattern a PCM patch. To avoid the damage to other component of the device, like the electrode wires, we need to use lithography to open a window, however, standard AIM-like flows encapsulate the photonic layer under 10 micrometres of SiO$_2$. Removing (or thinning) this cladding locally introduces topography steps and alignment challenges, as UV maskless exposure has alignment accuracy limitation and while ebeam lithography could provide high alignment accuracy, the top cladding, which is dielectric, prohibited the ebeam alignment due to the strong charging.

  \item \textbf{No straightforward electrical access.} The vertical junction in the microdisk biases the silicon core, not the PCM layer (Figure~\ref{fig:sfig45}(c)). Without redesigning the stack into a dedicated metal--PCM--metal (MIM) or metal--PCM--Si (MIS) heater, there is no direct electrical path to Joule-heat (or field-switch) the PCM. Relying solely on indirect substrate heating is inefficient and slow, and the thermal accumulation also damage the microdisk pn junction.
  
  \item \textbf{Thermal budget conflict.} Many chalcogenide PCMs (e.g., GST, Sb$_2$Se$_3$) require crystallization temperatures $>$\,150--250\,$^\circ$C and, in some cases, brief melt--quench cycles near 600\,$^\circ$C for re-amorphization. Such temperatures jeopardize doped junctions, metal stacks, and backend dielectrics that are rated for $\lesssim$\,300--350\,$^\circ$C.

\end{enumerate}

In summary, the combination of \emph{(i)} scarce and metal-covered surface area, \emph{(ii)} stringent thermal and contamination limits, \emph{(iii)} lack of a electrical probe path constraints makes conventional PCM trimming schemes poorly suited to this microdisk EOIM architecture.

\newpage
\section{Wire bonded chip}
\label{sec:s51}

Electrical wirebonding was performed by Optelligent, the whole chip has 256 metal wires that need wire bonding, from the wire-bonded chips ,we can clearly see the structure of the chip, which include the input edge coupler, Y-splitter, thermal phase shifter, microdisk modulator, 2D emitter array and 1D emitter array.

\begin{figure}[!htbp]
  \centering
  \includegraphics[width=0.6\textwidth]{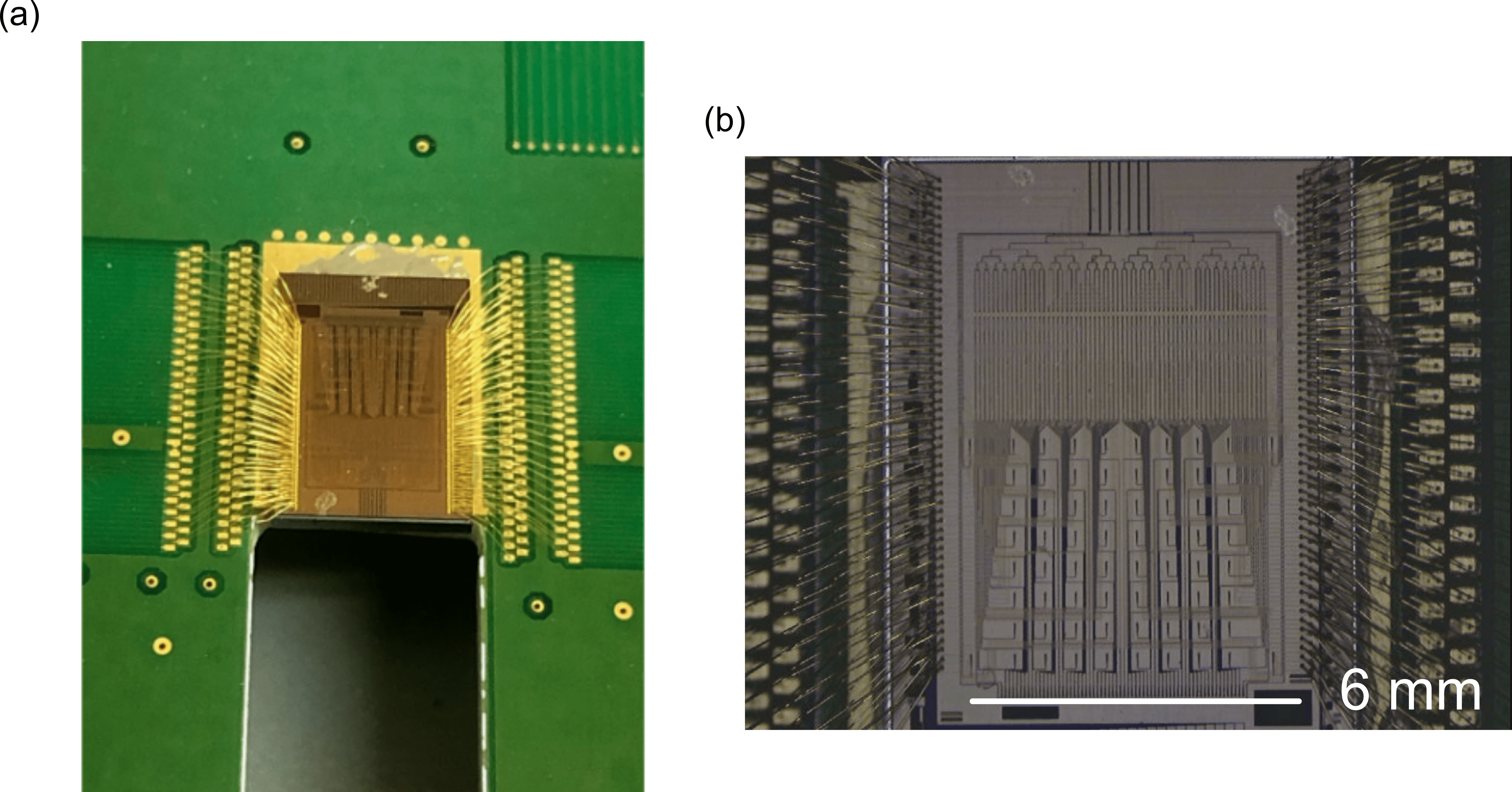}
  \caption{Schematic of the wire-bonded chip.}
  \label{fig:sfig51}
\end{figure}

\end{document}